\pdfoutput=1
\documentclass[10pt,pra,aps,twocolumn,showpacs,superscriptaddress]{revtex4-1}
\usepackage{latexsym}
\usepackage{graphics,epstopdf}
\usepackage{graphicx}
\usepackage[colorlinks=true]{hyperref}
\usepackage{float}
\usepackage{dcolumn}
\usepackage{bm}
\usepackage{hyperref}
\usepackage[mathlines]{lineno}
\usepackage{amsmath}
\usepackage{amsthm}
\usepackage{graphicx}
\usepackage{amsfonts}
\usepackage{amssymb}
\usepackage{makeidx}
\usepackage{float}
\usepackage[toc,page]{appendix}
\usepackage{natbib}
\theoremstyle{plain}

\newcount\colveccount
\newcommand*\colvec[1]{
        \global\colveccount#1
        \begin{pmatrix}
        \colvecnext
}
\def\colvecnext#1{
        #1
        \global\advance\colveccount-1
        \ifnum\colveccount>0
                \\
                \expandafter\colvecnext
        \else
                \end{pmatrix}
        \fi
}

\usepackage[left=2cm,right=2cm,top=3cm,bottom=4cm]{geometry}
  \newcommand{\pd}[2]{\frac{\partial #1}{\partial #2}}
  \newcommand{\bra}[1]{\left| #1 \right>} 
\newcommand{\ket}[1]{\left< #1 \right|} 
\newcommand{\braket}[2]{\left< #1 \vphantom{#2} \right|
 \left. #2 \vphantom{#1} \right>} 
 \newcommand{\matrixel}[3]{\left< #1 \vphantom{#2#3} \right| #2 \left| #3 \vphantom{#1#2} \right>} 

\begin{document}
\title{Manifestation of pointer state correlations in complex weak values of quantum observables}
\author{Som Kanjilal}
\email{somkanji@jcbose.ac.in}
\affiliation{Center for Astroparticle Physics and Space Science (CAPSS), Bose Institute, Kolkata-700091, India}
\author{Girish Muralidhara}
\email{girish.muralidhara@students.iiserpune.ac.in}
\affiliation{Indian Institute of Science Education and Research Pune, Pune-411 008 , India}
\author{Dipankar Home}
\email{dhome@jcbose.ac.in}
\affiliation{Center for Astroparticle Physics and Space Science (CAPSS), Bose Institute, Kolkata-700091, India}

\begin{abstract}
In the weak measurement (WM) scenario involving weak interaction and postselection by projective measurement, the empirical significance of weak values is manifested in terms of shifts in the measurement pointer's mean position and mean momentum. In this context, a general quantitative treatment is presented in this paper by taking into account the hitherto unexplored effect of correlations among the pointer degrees of freedom which pertain to an arbitrary multidimensional preselected pointer state. This leads to an extension of the earlier results, showing that, for complex weak values, the correlations among different pointer degrees of freedom can crucially affect the way the imaginary parts of the weak values are related to the observed shifts of the mean pointer position and momentum. The particular relevance of this analysis is discussed in the case of sequential weak interactions followed by a projective measurement enabling postselection (called sequential WM) which, in the special case, reduces to the usual WM scheme (involving a single weak interaction and postseletion) modified by the effect of pointer state correlations.
\end{abstract}
\maketitle


\section{Introduction}
		Formulation of the seminal idea of weak measurement (abbreviated, WM) in quantum mechanics by Aharanov, Albert and Vaidmann (AAV)\cite{aharonov} and its subsequent clarifications as well as elaborations \cite{legget, peres, aharonov1, Duck, aharonov2,vaid,tollak, brunner, WSG, haridass} over the years have given rise to a plethora of investigations, theoretical as well as experimental [for a recent comprehensive review see, for example, J.Dressel et al.\cite{dress}. This ranges from the use of WM in the analyses of intriguing quantum effects such as Hardy's paradox \cite{hardy}, the three box paradox\cite{3-box}, the quantum Cheshire Cat\cite{QCC}, to the application of WM in the context of quantum entanglement \cite{ent}, for verifying the ``error-disturbance uncertainty relations'' \cite{kane}, for observing the evolution of a quantum system in the semiclassical regime \cite{alexprl}, for avoiding loopholes in showing quantum violation of hybrid Bell-Leggett-Garg inequalities \cite{bell1}, for experimentally verifying Bell's inequality in time \cite{bell2}, for shedding light on quantum discord \cite{pati}, for demonstrating quantum contextuality \cite{pusey} and for studying tunneling \cite{stein1}arrival time \cite{arrival}, as well as for revealing interesting effects in the physics of telecommunication fibers \cite{tele}. Further, WM has been studied using neutron interferometry \cite{yuji} and has been invoked for high precision measurements concerning quantum metrology \cite{metro} such as for identifying a tiny spin Hall effect\cite{spinhall}, for detecting very small transverse beam deflections \cite{TBD,Goswami:14}, and tiny temporal delay \cite{TS}. Interestingly, it has also been used for directly measuring the quantum wave function \cite{QWF} and for discerning signatures of the average quantum trajectories for photons \cite{stein}. Against this backdrop, in order to motivate our work, it will be useful to recapitulate the essence of the standard WM scenario.\\
		
Let us consider the preparation of a given system in an appropriate preselected state $\bra{\psi_{i}}$, with the state being subjected to a von Neumann type interaction described by the Hamiltonian $\hat{H}=g(t)\hat{A} \otimes \hat{q}$ where $\hat{A}$ is the system observable, $\hat{q}$ is the measurement pointer observable and $g(t)$ is the coupling parameter given by the normalized compact support around the time of measurement \citep{Duck}; here the von Neumann coupling is assumed to be `weak' in the following sense. Taking the initial pointer state to be, say, a Gaussian wave function, the interaction involved is said to be `weak' if it results in the pointer state to be a superposition of Gaussian wave functions which are substantially overlapping. Subsequently, this interaction is followed by an appropriate postselection pertaining to the projective measurement of any system observable other than $\hat{A}$ which is involved in the preceding weak interaction. Then the superposition of overlapping Gaussian wave functions, in effect, gives rise to a single slightly shifted Gaussian wave function \cite{Duck, WSG}. The net effect is manifested in terms of the shifted probability distribution of the pointer variable corresponding to the postselected system state $|\psi_f \rangle$ which is an eigenstate of one of the outcomes of the projective measurement in question. The key result shown by AAV is that the weak interaction involving the system observable $\hat{A}$, combined with postselection, results in the final shift of the postselected pointer variable distribution that turns out to be proportional to a quantity called the `weak value' of the observable $\hat{A}$, which is defined as\\
\begin{equation}
\label{e1}
(A)_{w}=\frac{\langle\psi_{f}|\hat{A}|\psi_{i}\rangle}{\langle\psi_{f}|\psi_{i}\rangle}
\end{equation}      
Note that the weak value $(A)_{w}$ is, in general, a complex quantity. Its empirical significance was first pointed out in the footnote 4 of the paper by Aharonov, Albert and Vaidman \cite{aharonov} as well as in the paper by Aharonov and Vaidman \cite{vaid} using the Gaussian function for the pointer state. later, the more general quantitative relations linking the shifts of the mean pointer position and momentum with the real and imaginary parts of the weak value were formulated by Jozsa \cite{jozsa}. To put it precisely in the context of weak interaction involving the Hamiltonian $H=g(t)\hat{A}\otimes \hat{q}$, the final and the initial expectation values of pointer position ($\langle \hat{q} \rangle_{f},\langle \hat{q} \rangle_{i}$)and pointer momentum ($\langle \hat{p} \rangle_{f},\langle \hat{p} \rangle_{i}$) are respectively mutually related as follows 
\begin{equation}
\label{sqm2}
\langle \hat{q} \rangle_{f}=\langle \hat{q} \rangle_{i}+2 \lambda ({Im}(A)_{w}) var(q)
\end{equation} 
\begin{equation}
\label{sqm3}
\langle \hat{p} \rangle_{f}=\langle \hat{p} \rangle_{in} - \lambda\text{Re} (A)_{w} + m\lambda (\text{Im} (A)_{w})\pd{var(q)}{t}
\end{equation} 
where the variance $var(q)=\langle \hat{q}^{2}\rangle_{i}-\langle \hat{q} \rangle_{i}^{2}$, $\int g(t)dt= \lambda$, and m is the mass of the system in question. It is the above results that constitute the specific starting point of this work.\\

Before proceeding further, an important basic point to stress is that since a typical WM scheme uses at-least two von Neumann interactions involving, in general, two different pointer degrees of freedom (one for weak interaction and the other for projective measurement leading to the postselection), the scenario in general involves the use of multidimensional pointer variable distribution. \\

Here it is relevant to emphasize that, given any covariance matrix, it is possible to generate a multivariate distribution embodying the correlations given by the covariance matrix (Cholesky decomposition). Now, note that while the effects of multivariate pointer state distribution without correlations have already been discussed in the context of weak measurement \cite{alexprl,stein3,prac,cum,prac1}, curiously, the possible effects of multidimensionality embodying \textit{correlations} among different pointer degrees of freedom have remained largely unexplored. This holds apart from a couple of works probing the effects of multidimensionality of pointer states in the special case of two dimensional Hermite-Gaussian and Laguerre-Gaussian optical modes as pointer states \cite{JQV2}. Against this backdrop, in this paper we seek to provide a hitherto unexplored general framework for treating the effects of correlations among different pointer degrees of freedom, which has a special significance, for example, in the context of continuous variable entanglement as discussed in the final section of this paper.\\

Note that, in the usual treatments, including in Jozsa's derivation of Eqs. (\ref{sqm2}) and (\ref{sqm3})\cite{jozsa}, the underpinning assumption is either that the multidimensional pointer wave function is factorizable, or that the same pointer degree of freedom which is weakly coupled to the system degree of freedom is used for projective measurement enabling postselection. A key aspect of our treatment is that by taking into account a  general preselected multidimensional pointer state, we consider the case where the pointer degree of freedom involved in the final projective measurement is \textit{different} from that used in the preceding weak interaction or weak interactions in the case of sequential WM. The key feature arising from the latter aspect, as shown in our paper for the case of sequential WM(Sec. II), as well as for the usual WM scenario(Sec. III), is that if the \textit{initial pointer state involves correlations between the pointer degrees of freedom which are involved in weak interaction(s) and postselection} , the final expectation values of the pointer degrees of freedom  will contain contributions from these non-zero correlations, depending upon whether the relevant non-vanishing weak value(s) are complex or not. The essential result demonstrated is the way the quantitative relations between mean pointer position (momentum) and weak values given by Eqs. (\ref{sqm2}) and (\ref{sqm3}) get significantly modified in the presence of correlations between different pointer degrees of freedom. \\

Here we may observe that a noteworthy work using the initial pointer wave function as a multidimensional function is that by G. Mitchison \cite{cum}. In this work he considered the joint expectation values of postselected pointer degrees of freedom in terms of joint weak values and generalized the treatment for weak measurement involving arbitrary number of weak interactions. However, the multidimensional initial pointer wave function considered by Mitchison is factorizable. On the other hand, in this paper, we consider essentially the effect of nonfactorizability embodying correlations in the initial pointer wave function including \textit{all} the pointer degrees of freedom. Another interesting line of works using multidimensional pointer wave function involves extracting joint weak value involving a product of two single particle operators\cite{stein3} and subsequently extending it for joint weak values of the product of N single particle operators \cite{prac}. Again, these studies also essentially use factorizable initial pointer wave function and hence the possible effect of correlations in the multidimensional pointer wave function remains unanalyzed.\\

Now, for outlining our scheme, considering sequential WM, we use a weak interaction involving a system variable ($\hat{A_{1}}$) coupled with a pointer degree of freedom ($\hat{q}_{1}$ or $\hat{p}_{1}$), followed by another weak interaction involving a system variable ($\hat{A_{2}}$) coupled with a pointer degree of freedom ($\hat{q}_{2}$ or $\hat{p}_{2}$). It is then found that the postselection using a projective measurement involving the pointer variable ($\hat{q}_{3}$ or $\hat{p}_{3}$) results in the individual shifts along different axes (denoted by the lower index i = 1, 2, 3) of the postselected three dimensional pointer variable distribution. Each of these shifts has contributions from \textit{both} the weak values ($( \hat{A_{1}})_{w}$ and $(\hat{A_{2}})_{w}$), arising from the two successive weak interactions considered, apart from being dependent on the \textit{correlations} between the pointer degrees of freedom. To compute these shifts, we evaluate the final expectation values of the respective degrees of freedom ($\langle \hat{q}_{1,2,3} \rangle_{f}$ or $\langle \hat{p}_{1,2,3} \rangle_{f}$) pertaining to the postselected pointer state, relating each of these expectation values to both the weak values ($(A_{1})_{w}$ and $(A_{2})_{w}$), along with the correlation terms involving the pointer degrees of freedom. For this demonstration, it suffices to consider the strength of each weak interaction up to first order. An important significance of the aforementioned extension is that the correlation terms appear on the final shifts of the pointer degrees of freedom only when the imaginary parts of the respective weak values are \textit{nonzero}.Here we may stress that , as in Jozsa's result, we do not consider the effect of the time evolution of the probe state that may occur before detection, which has been taken into account by Lorenzo and Egues \cite{lorenzo}\\

The archetypal investigation to date concerning the sequential WM by Mitchison et al.\cite{SQM},while considering the strength of each weak interaction up to second order, has essentially calculated the expectation value of a product of pointer variables pertaining to the post-selected pointer state \textit{without} taking into account the possible effects of correlations among the pointer degrees of freedom. To be precise, Mitchison et al.\cite{SQM} obtained the joint expectation value of $ \hat{q}_{1} \hat{q}_{2}$ in terms of the joint weak value $( A_{1}A_{2} )_{w}$ and the product of the individual weak values $(A_{1})_w$ and $(A_{2})_w$. The fundamental difference between their work and ours as discussed in this paper is, thus, easily evident.\\
 
In the following Section II we proceed to delineate the mathematical details of our treatment in the case of sequential WM showing explicitly the way the empirical signature of weak values in terms of the observed shifts along different axes of the pointer variable distribution involves correlations among the pointer degrees of freedom. Then, in Section III, we discuss how in the special case of the vanishing strength of one of the two weak interactions, i.e., in the usually considered WM scenario, our treatment reveals the effects of the pointer state correlations on the individual shifts along different axes pertaining to the multidimensional pointer variable distribution, thereby affecting their relation with the weak value in question. In the concluding Section IV, we indicate some implications of this work as well as a few directions for further studies, including possibilities of future empirical probing. 

\section{The treatment of sequential WM in the presence of correlations between pointer degrees of freedom}    
Let the initial joint state of the system and the measurement pointer be given by
\begin{equation}
\label{sqm4}
\bra{\Psi}=\bra{\psi_{i}}\otimes\bra{\phi_{i}}
\end{equation}
where $|\psi_{i}\rangle$ and $|\phi_{in}\rangle$ are respectively the system and the pointer initial states that are taken to be three dimensional.

If $\phi_{i}(q_{1},q_{2},q_{3}) \neq \phi_{i}(q_{1})\phi_{i}(q_{2})\phi_{i}(q_{3})$, this would imply that the pointer state is \textit{not} correlated between any two of the $3$ pointer degrees of freedom, i.e., the relevant covariance matrix \cite{Kampen20071} whose elements are given by $\Sigma_{ij}=\langle (\hat{q_{i}}-\langle \hat{q_{i}}\rangle)(\hat{q_{j}}-\langle \hat{q_{j}}\rangle)\rangle$ is diagonal, meaning that only the variance terms of this matrix are non-vanishing. In general, the covariance matrix is, however, not diagonal, with the non vanishing off-diagonal terms, i.e., the correlation terms with respect to the initial pointer state being given by  
\begin{widetext}
\begin{equation}
\label{e5}
corr(q_{l},q_{m})_{i}=\int \phi_{i}^{*}(p_{1},p_{2},p_{3})\hat{q}_{l}\hat{q}_{m}\phi_{i}(p_{1},p_{2},p_{3})d\vec{p}-\int \phi_{i}^{*}(p_{1},p_{2},p_{3})\hat{q}_{l}\phi_{i}(p_{1},p_{2},p_{3})d\vec{p}\int \phi_{i}^{*}(p_{1},p_{2},p_{3})\hat{q}_{m}\phi_{i}(p_{1},p_{2},p_{3})d\vec{p} \hspace{5mm}\text{$l \neq m$}
\end{equation}
\end{widetext}

where $d\vec{p}=dp_{1}dp_{2}dp_{3}$. Before proceeding further, we note the following properties of the correlation terms. Consider the correlation function of momentum displaced initial pointer wave function 
$\phi_{i}(p_{1}-l_{1},p_{2}-l_{2},p_{3}-l_{3})$ ($l_{i}$ is the amount of displacement of momentum $p_{i}$), given by
\begin{widetext}
\begin{align}
corr(p_{l},p_{m})_{i,l_{1},l_{2},l_{3}} & = \int \phi_{i}^{*}(p_{1}-l_{1},p_{2}-l_{2},p_{3}-l_{3})p_{l}p_{m}\phi_{i}(p_{1}-l_{1},p_{2}-l_{2},p_{3}-l_{3})d\vec{p}-\int \phi_{i}^{*}(p_{1}-l_{1},p_{2}-l_{2},p_{3}-l_{3})p_{l}\nonumber \\
\label{x1}
& \phi_{i}(p_{1}-l_{1},p_{2}-l_{2},p_{3}-l_{3})d\vec{p}\int \phi_{i}^{*}(p_{1}-l_{1},p_{2}-l_{2},p_{3}-l_{3})p_{m}\phi_{i}(p_{1}-l_{1},p_{2}-l_{2},p_{3}-l_{3})d\vec{p} \\
& = \int \phi_{i}^{*}(p_{1},p_{2},p_{3})(p_{l}+l_{l})(p_{m}+l_{m})\phi_{i}(p_{1},p_{2},p_{3})d\vec{p}- \int \phi_{i}^{*}(p_{1},p_{2},p_{3})(p_{l}+l_{l})\phi_{i}(p_{1},p_{2},p_{3})d\vec{p}\nonumber\\
& \int \phi_{i}^{*}(p_{1},p_{2},p_{3})(p_{m}+l_{m})\phi_{i}(p_{1},p_{2},p_{3})d\vec{p} \nonumber\\
\label{x3}
& = corr(p_{l},p_{m})_{i}
\end{align}
\end{widetext}
 Similarly, it can be shown that
 \begin{equation}
 \label{x4}
 corr(q_{l},p_{m})_{i,l_{1},l_{2},l_{3}} = corr(q_{l},p_{m})_{i} 
 \end{equation}
 \begin{equation}
 \label{x5}
 corr(q_{l},q_{m})_{i,l_{1},l_{2},l_{3}} = corr(q_{l},q_{m})_{i} 
 \end{equation}
The above equations signify that the correlation functions are not dependent on the displacement of the momentum distribution of the pointer wave function. Note that, $corr(q_{l},q_{l})=var(q_{l})$ and $corr(p_{l},p_{l})=var(p_{l})$. Therefore, in the light of above relations it can be stated that
 \begin{equation}
 \label{x7}
 var(q_{l})_{i,l_{1},l_{2},l_{3}}=var(q_{l})_{i}
 \end{equation} 
 \begin{equation}
 \label{x8}
 var(p_{l})_{i,l_{1},l_{2},l_{3}}=var(p_{l})_{i}
 \end{equation}
 
In our treatment, for generality, we take the initial pointer state to be as follows
\begin{equation}
\label{sqm5}
\bra{\phi_{i}(p_{1},p_{2},p_{3})}=\int \phi_{i}(p_{1},p_{2},p_{3})|p_{1}\rangle|p_{2}\rangle |p_{3}\rangle d\vec{p}
\end{equation} 
 If $\phi_{i}(p_{1},p_{2},p_{3})$ involves correlations, then the question addressed in this paper is whether and, if so, how the shifts of the relevant pointer degrees of freedom will capture the effect of these correlations.

The successive weak interactions in the setup considered here are taken to be of the von Neumann type weak coupling between the system and the pointer observables, where the two Hamiltonians in question are $H_{1}= g_{1}(t)\hat{A}_{1}\otimes\hat{q_{1}}$ and $H_{2}= g_{2}(t)\hat{A}_{2}\otimes\hat{q_{2}}$. We assume that, apart from the von Neumann couplings used, the system evolves freely in-between the weak interactions and before being subjected to the postselection.
The postselection is performed by using projective measurement involving von Neumann type strong coupling between $\hat{A}_{3}$ and $\hat{q}_{3}$, (the corresponding Hamiltonian $H=g_{t}\Sigma_{k}a_{3k}\hat{\Pi}_{3k}\otimes\hat{q}_{3}$ with $\int g_{3}(t)dt=1$, where $\hat{\Pi}_{3k}=\bra{a_{3k}}\ket{a_{3k}}$ is the projection operator corresponding to $\bra{a_{3k}}$ with eigenvalue $a_{3k}$). Taking $\int g_{2}(t)dt = \lambda_{2}$ and $\int g_{1}(t)dt = \lambda_{1}$, if we expand both the exponentials occurring in the evolution operators of $H_{1}$ and $H_{2}$ up to first order of $\lambda_{1}$ and $\lambda_{2}$ respectively, the joint state of system and pointer after the strong von Neumann interaction but before postselection can be written as follows
\begin{align}
\bra{\psi}& = e^{-i\Sigma_{k}a_{3k}\hat{\Pi}_{3k}\otimes\hat{q}_{3}}\int d\vec{p}(1+i\lambda_{1}\hat{A}_{1}\otimes\hat{q}_{1}+i\lambda_{2}\hat{A}_{1}\otimes\hat{q}_{2}) \nonumber\\
\label{ss3}
& \phi_{i}(p_{1},p_{2},p_{3})\bra{\psi_{i}}\bra{p_{1}}\bra{p_{2}}\bra{p_{3}}
\end{align}
Using $e^{i\Sigma_{k}a_{3k}\hat{\Pi}_{3k}\otimes\hat{q}_{3}}=\Sigma_{k}e^{ia_{3k}\hat{q}_{3}}\Pi_{3k}$, the expression for the weak value given by Eq.(\ref{e1}) and $e^{ia_{3k}\hat{q}_{3}}\phi_{i}(p_{1},p_{2},p_{3})=\phi_{i}(p_{1},p_{2},p_{3}-a_{3k})$, Eq. (\ref{ss3}) can be modified as follows
\begin{align}
\bra{\psi} & = \Sigma_{k}\braket{a_{3k}}{\psi_{i}}\int dp_{1}dp_{2}dp_{3}(1+i\lambda_{1}(A_{1})_{w}\otimes\hat{q_{1}}+\nonumber \\ 
\label{ss2}
& i\lambda_{2}(A_{2})\otimes\hat{q_{2}})\phi_{i}(p_{1},p_{2},p_{3}-a_{3k})\bra{a_{3k}}\bra{p_{1}}\bra{p_{2}}\bra{p_{3}} 
\end{align}
It is to be noted that, after successive weak interactions, the pointer degrees of freedom $q_{1}$ and $q_{2}$ get entangled with the system observables $A_{1}$ and $A_{2}$. Presence of nonzero correlation between $q_{1,2}$ and $q_{3}$ and/or $p_{3}$ would imply that before the strong von Neumann interaction, the system observables $A_{1}$ and $A_{2}$ are further correlated with $q_{3}$ and/or $p_{3}$. Subsequently, it can be seen from Eq. (\ref{ss2}) that the strong von Neumann interaction creates an entanglement between $\bra{a_{3k}}$ and $\bra{p_{3}}$.\\

 Postselecting the system state onto $|a_{3l}\rangle$ which is one of the eigenstates of the system variable $\hat{A_{3}}$ one can obtain the relevant pointer state as follows(taking $\hbar=1$ throughout our treatment)\\ 
	  \begin{align}
	  |\phi_{f,3l}\rangle & \approx \braket{a_{3l}}{\psi_{i}}\int (1+i\lambda_{2}(A_{2})_{w}\hat{q}_{2}+i\lambda_{1}(A_{1})_{w}\hat{q}_{1})\rangle \nonumber\\
	  \label{sqm8}
	  & \phi_{i}(p_{1},p_{2},p_{3}-a_{3l})\bra{p_{1}}|p_{2}\rangle\bra{p_{3}}dp_{1}dp_{2}dp_{3}
	  \end{align}
	    Let us consider the final expectation value of an arbitrary pointer variable $M$, corresponding to the postselected pointer state $|\phi_{f,3l}\rangle$ given by
	  \begin{equation}
	  \label{sqm9}
	  \langle \hat{M} \rangle_{f,3l} = \frac{\matrixel{\phi_{f,3l}}{\hat{M}}{\phi_{f,3l}}}{\braket{\phi_{f,3l}}{\phi_{f,3l}}}
	  \end{equation}
	  Then, writing the relevant weak values occurring in Eq. (\ref{sqm8}) as follows
	  \begin{equation}
	  \label{e11}
	  (A_{1})_{w}=a_{1}+ib_{1}
	  \end{equation}
	  \begin{equation}
	  \label{e12}
	  	  (A_{2})_{w}=a_{2}+ib_{2}
	  	  \end{equation}
	  	  and, using Eq. (\ref{sqm8}), the numerator and the denominator of Eq. (\ref{sqm9}) are respectively given by
	 
	 \begin{align}
	 \matrixel{\phi_{f}}{\hat{M}}{\phi_{f,3l}} & \approx |\braket{a_{3l}}{\psi_{i}}|^{2}(\langle \hat{M} \rangle_{i,3l}+i\lambda_{1}a_{1}\langle [\hat{M},\hat{q_{1}}] \rangle_{i,3l}\nonumber \\
	 &  +i\lambda_{2}a_{2}\langle [\hat{M},q_{2}] \rangle_{i,3l}-\lambda_{1}b_{1}\langle \{\hat{M},q_{1}\} \rangle_{i,3l} \nonumber \\
	 \label{sqm13}
	 & -\lambda_{2}b_{2}\langle \{\hat{M},\hat{q_{2}}\} \rangle_{i,3l})
	 \end{align}
	 where,
\begin{equation}
\label{x10}
\langle A \rangle_{i,3l}=\int \phi_{i}^{*}(p_{1},p_{2},p_{3}-a_{3l})\hat{A}\phi_{i}(p_{1},p_{2},p_{3}-a_{3l})dp_{1}dp_{2}dp_{3}
\end{equation}	 
Similarly
	 
	  \begin{equation}
	 \label{sqm12}
	 \braket{\phi_{f,3l}}{\phi_{f,3l}} \approx |\braket{a_{3l}}{\psi_{i}}|^{2}(1-2\lambda_{1}b_{1}\langle \hat{q_{1}}\rangle_{i,3l}-2\lambda_{2}b_{2}\langle \hat{q_{2}}\rangle_{i,3l})
	 \end{equation}
	 
	 Next, using Eqs. (\ref{sqm13}) and (\ref{sqm12}), from Eq. (\ref{sqm9}) one can obtain the value of $\langle \hat{M} \rangle_{f,3l}$ up to the first order in $\lambda_{1},\lambda_{2}$ given by 
	 \begin{align}
	 \langle \hat{M} \rangle_{f,3l} & \approx(\langle \hat{M} \rangle_{i,3l}+i\lambda_{1}a_{1}\langle [\hat{M},\hat{q_{1}}] \rangle_{i,3l}+i\lambda_{2}a_{2}\langle [\hat{M},\hat{q_{2}}] \rangle_{i,3l} \nonumber \\
	 & -\lambda_{1}b_{1}\langle \{\hat{M},\hat{q_{1}}\} \rangle_{i,3l}-\lambda_{2}b_{2}\langle \{\hat{M},\hat{q_{2}}\} \rangle_{i,3l})\nonumber \\
	\label{sqm14} 
	 & (1+2\lambda_{1}b_{1}\langle \hat{q_{1}}\rangle_{i,3l}+2\lambda_{2}b_{2}\langle \hat{q_{2}}\rangle_{in,3l})
	 \end{align}
	 Here $[...]$ and $\{....\}$ denote respectively the commutator and the anti-commutator. For the specific choice of the pointer observable $\hat{M}=\hat{q}_{1}$ in Eq. (\ref{sqm14}), the relevant commutators vanish and one obtains
	      \begin{align}
	      \label{sqm16.5}
	      \langle \hat{q_{1}} \rangle_{f,3l} & = \langle \hat{q_{1}} \rangle_{i,3l}-2\lambda_{1}Im(A_{1})_{w}var(q_{1})_{i,3l}\nonumber \\
	      & -2\lambda_{2}Im(A_{2})_{w}corr(q_{1},q_{2})_{i,3l}
	         \end{align}
	         Using relevant forms of Eqs. (\ref{x5}) and (\ref{x7}) we can rewrite Eqn. (\ref{sqm16.5}) as follows 
	         \begin{align}
	      \label{sqm16}
	      \langle \hat{q_{1}} \rangle_{f,3l} & = \langle \hat{q_{1}} \rangle_{i}-2\lambda_{1}Im(A_{1})_{w}var(q_{1})_{i}\nonumber \\
	      & -2\lambda_{2}Im(A_{2})_{w}corr(q_{1},q_{2})_{i}
	         \end{align}
	         For $M=q_{2}$ we will obtain
	     \begin{align}
	          \label{sqm18}
	          \langle \hat{q}_{2} \rangle_{f,3l}=\langle \hat{q_{2}} \rangle_{i}-2\lambda_{2}Im(A_{2})_{w}var(q_{2})_{i}\nonumber \\
	          -2\lambda_{1}Im(A_{1})_{w}corr(q_{1},q_{2})_{i}
	       \end{align}
	       using Eqs.(\ref{e11}) and (\ref{e12}), and where the correlation term $corr(q_{1}, q_{2})_{i}$ is given by Eq. (\ref{e5}) for $l, m = 1, 2$.
	     
	             The key consequence of the presence of the correlation term $corr(q_{1},q_{2})$ in the above equations can be expressed as follows. Eq. \ref{sqm16}(\ref{sqm18}) shows that the shift of the expectation value of the pointer degree of freedom $q_{1}(q_{2})$, apart from being dependent on the imaginary part of the weak value of the system observable $\hat{A_{1}}(\hat{A_{2}})$ which is coupled (\textit{a la} von Neumann) with the pointer degree of freedom $q_{1}(q_{2})$, contains an additional contribution arising from the correlation term $corr(q_{1},q_{2})_{i}$ that depends on the imaginary part of the weak value of the system observable $\hat{A_{2}}(\hat{A_{1}})$ which is, too, von Neumann coupled with the pointer degree of freedom $\hat{q}_{2}(\hat{q}_{1})$.  
	  
	         For $\hat{M}=\hat{q}_{3}$, one obtains from Eq. (\ref{sqm14})
	         \begin{align}
	         \label{sqm18.55}
	         \langle \hat{q}_{3} \rangle_{f,3l}& = \langle \hat{q}_{3} \rangle_{i,3l}-2\lambda_{1}Im(A_{1})_{w}corr(q_{1},q_{3})_{i,3l}\nonumber \\
	         & -2\lambda_{2}Im(A_{2})_{w}corr(q_{2},q_{3})_{i,3l}
	         \end{align}
	         Using $\langle \hat{q}_{3} \rangle_{i,3l}=\langle \hat{q}_{3} \rangle_{i}$  and Eq. (\ref{x5}) we can recast Eq. (\ref{sqm18.55}) as follows
	         \begin{align}
	         \label{sqm18.5}
	         \langle \hat{q}_{3} \rangle_{f,3l}& = \langle \hat{q}_{3} \rangle_{i}-2\lambda_{1}Im(A_{1})_{w}corr(q_{1},q_{3})_{i}\nonumber \\
	         & -2\lambda_{2}Im(A_{2})_{w}corr(q_{2},q_{3})_{i}
	         \end{align}
	         
	         Note that the shift in the expectation value of the pointer degree of freedom $q_{3}$ arising from sequential weak interactions cum postselction essentially depends upon the  non-zero correlations between $q_{1,2}$ and $q_{3}$ present in the initial pointer wave function. Presence of these correlations ensure that after two successive weak interactions, not only $q_{1,2}$ but also $q_{3}$ gets entangled with the system observables $A_{1}$ and $A_{2}$. It is this entanglement due to which the postselection of a particular system state ensures that the expectation value of $q_{3}$ gets shifted.\\

	         Considering the sequential WM scenario, if the pointer degree of freedom involved in the projective measurement enabling postselection is the \textit{same} as that used in the preceding weak interactions, then Eq. $(\ref{sqm18.5})$ reduces to
	         \begin{equation}
	         \label{sq1}
	         \langle \hat{q}_{3} \rangle_{f}=\langle \hat{q}_{3} \rangle_{i}-2\lambda_{1}Im(A_{1})_{w}var(q_{3})-2\lambda_{2}Im(A_{2})_{w}var(q_{3})
	         \end{equation}
	          which does not contain any effect of correlations between the pointer degrees of freedom present in the initial state.\\
	          
	          Now, if the pointer degrees of freedom that are involved in the weak interactions are \textit{different} from that used in postselection, then the shift of the final expectation value of the pointer degree of freedom ($q_{3}$) involved in postselection will contain the effect of correlations embodied in the initial preselected pointer state. When these correlations vanish, such a shift will also vanish.
	          It is the above feature that is reflected in Eq.(\ref{sqm18.5}) which essentially pertains to the case where the pointer degree of freedom $q_{3}$ used in postselection is different from the pointer degrees of freedom ($q_{1},q_{2}$) used in the preceding weak interactions. In such cases, the shift of the final expectation value of $q_{3}$ crucially depends on whether at least one of the correlations $corr(q_{1},q_{3})$ or $corr(q_{2},q_{3})$ is non-vanishing.\\
	         
	         Further, note that, even if all these correlations vanish, the shifts of the final expectation values of the pointer degrees of freedom $q_{1},q_{2}$ occurring in the weak interactions remain non-vanishing, as can be seen from Eqs. (\ref{sqm16}) and (\ref{sqm18}).\\
	         
        Similarly, for $\hat{M}=\hat{p}_{1,2}$, from Eq. (\ref{sqm14}) we get       
        \begin{align}
        \langle \hat{p}_{1,2} \rangle_{f,3l} & \approx(\langle \hat{p}_{1,2} \rangle_{i,3l}+i\lambda_{1}a_{1}\langle 
       [\hat{p}_{1,2},\hat{q}_{1}] \rangle_{i,3l}+i\lambda_{2}a_{2}\langle [\hat{p}_{1,2},\hat{q}_{2}] \rangle_{i,3l} \nonumber \\
	 & -\lambda_{1}b_{1}\langle \{\hat{p}_{1,2},\hat{q}_{1}\} \rangle_{i,3l}-\lambda_{2}b_{2}\langle \{\hat{p}_{1,2},\hat{q}_{2}\} \rangle_{i,3l})  \nonumber\\ 
	 \label{sqm18.75}
	 & (1+2\lambda_{1}b_{1}\langle \hat{q}_{1}\rangle_{i,3l}+2\lambda_{2}b_{2}\langle \hat{q}_{2}\rangle_{i,3l}) 
	 \end{align} 
	   Using the canonical commutation relations $[\hat{q}_{1},\hat{p}_{2}]=0,[\hat{q}_{1},\hat{p}_{1}]=i\mathbb{I},[\hat{q}_{2},\hat{p}_{1}]=0,[\hat{q}_{2},\hat{p}_{2}]=i\mathbb{I}$, Eqs. (\ref{x4}),(\ref{x7}) and following the  mathematical treatment of Jozsa \cite{jozsa} we can obtain
	 \begin{align}
	 \label{sqm19}
	 \langle \hat{p}_{1} \rangle_{f,3l} & = \langle \hat{p}_{1} \rangle_{i}+\lambda_{1}Re(A_{1})_{w}+ m\lambda_{1}Im(A_{1})_{w}\pd{var(q_{1})_{i}}{t}\nonumber \\
	 & +2\lambda_{2}Im(A_{2})_{w}corr(p_{1},q_{2})_{i}
	 \end{align}
     \begin{align}
	 \label{sqm20}
	 \langle \hat{p}_{2} \rangle_{f,3l}& = \langle \hat{p}_{2} \rangle_{i}+\lambda_{2}Re(A_{2})_{w}+m\lambda_{2}Im(A_{2})_{w}\pd{var(q_{2})_{i}}{t}\nonumber \\ 
	 & +2\lambda_{1}Im(A_{1})_{w}corr(q_{1},p_{2})_{i}
	 \end{align}\\
	 where $corr(q_{l},p_{m})_{i}$ with $i \neq j$ is given by
	 \begin{equation}
	 \label{sqm19.5}
	 corr(q_{l},p_{m})_{i}=\langle \hat{q}_{l}\hat{p}_{m} \rangle_{i} - \langle \hat{q}_{l} \rangle_{i}\langle \hat{p}_{m} \rangle_{i}
	 \end{equation}
 Here again, the effect of the pointer state correlation is embodied in Eq. \ref{sqm19}(\ref{sqm20}) in terms of the shift of the expectation value of the pointer degree of freedom $p_{1}(p_{2})$ containing an additional contribution from the imaginary part of the weak value of the system observable $A_{2}(A_{1})$ which is coupled (\textit{a la} von Neumann) with  the pointer degree of freedom $q_{2}(q_{1})$.

  For $\hat{M}=\hat{p}_{3}$, from Eq. (\ref{sqm14}) one can similarly obtain
   	 \begin{align}
 	 \label{sqm21.55}
 	  \langle \hat{p}_{3} \rangle_{f,3l} & = \langle \hat{p}_{3} \rangle_{i,3l}+2\lambda_{1}Im(A_{1})_{w}corr(q_{1},p_{3})_{i,3l}\nonumber \\
 	  & +2\lambda_{2}Im(A_{2})_{w}corr(q_{2},p_{3})_{i,3l}
 	 \end{align}
 	 Note that $\langle \hat{p}_{3} \rangle_{i,3l}=\langle \hat{p}_{3} \rangle_{i}+a_{3l}$. Therefore, using Eq. (\ref{x4}) we obtain
 	 \begin{align}
 	 \label{sqm21}
 	  \langle \hat{p}_{3} \rangle_{f,3l} & = a_{3l}+\langle \hat{p}_{3} \rangle_{i}+2\lambda_{1}Im(A_{1})_{w}corr(q_{1},p_{3})_{i}\nonumber \\
 	  & +2\lambda_{2}Im(A_{2})_{w}corr(q_{2},p_{3})_{i}
 	 \end{align}
 	 In the absence of weak interactions, the final expectation value of $p_{3}$ will get shifted by the amount $a_{3l}$ due to projective measurement. In the presence of successive weak interactions involving $\hat{q}_{1}$ and $\hat{q}_{2}$ pointer degrees of freedom, entanglement between $p_{3}$ and the system observables $A_{1}$ and $A_{2}$ is created through non-zero values of $corr(q_{1}p_{3})$ and $corr(q_{2},p_{3})$- this results in the further shift of the expectation value of $p_{3}$ which can be seen from Eq. (\ref{sqm21}).
 	 
 	 It may be stressed here that the correlation terms among the pointer position and momenta degrees of freedom occurring in Eqs. (\ref{sqm19}) - (\ref{sqm21}) are non-vanishing essentially because of the non-vanishing correlation among the position degrees of freedom occurring in the preselected pointer state. This can be seen by recalling that the Fourier transform of a multidimensional probability density function, say, $f(q_{1}, q_{2}, q_{3})$ can be done through intermediate steps, each step comprising Fourier transform of $q_{i}$ to $p_{i}$ which ensures that if $corr(q_{i}, q_{j}) \neq 0$, then $corr(q_{i}, p_{j})$ is also necessarily non-vanishing (see Appendix A for the relevant mathematical details).
 	 
 	 It is the results given by Eqs. (\ref{sqm16}) - (\ref{sqm18.5}) and (\ref{sqm19}) - (\ref{sqm21}) which provide the extension of Jozsa's results (given by Eq. (\ref{sqm2}) and (\ref{sqm3})) for the sequential WM in the presence of pointer state correlations. Note that, if all the correlation terms vanish in Eqs. (\ref{sqm16}) - (\ref{sqm18.5}) and (\ref{sqm19}) - (\ref{sqm21}), i.e., if the three-dimensional pointer state is separable in different pointer degrees of freedom, then Eqs. (\ref{sqm16}) - (\ref{sqm18.5}) and (\ref{sqm19}) - (\ref{sqm21}) reduce to Jozsa's results in the three dimensional case. Thus, an upshot of our analysis is that if the strengths of both the weak interactions are taken to be up to first order, Jozsa's results (although originally derived for the WM scenario using single weak interaction) remain valid for the case of sequential WM, too, provided the pointer state correlations are ignored.
 	     
 \section{The treatment of WM with single weak interaction in the presence of pointer state correlations}
In the case of the usually considered WM scenario involving a single weak interaction, one needs essentially two pointer degrees of freedom, one for the von Neumann weak interaction and the other for implementing postselection via a projective measurement. Thus, in this context, it suffices to consider the preselected pointer state to be of the two-dimensional form 
\begin{equation}
\label{sqm22}
\bra{\phi_{i}(p_{1},p_{2})}=\int \phi_{i}(p_{1},p_{2})\bra{p_{1}}\bra{p_{2}}d\vec{p}
\end{equation} 
where we take $corr(q_{1},q_{2})_{i} \neq 0$.\\

Here again the preselected system state is $\bra{\psi_{i}}$, while the von Neumann coupling for weak interaction is of the usual form $H= \lambda_{1}\hat{A}_{1}\otimes\hat{q}_{1}$ and the subsequent postselection is done through projective measurement using another von Neumann coupling involving the  pointer degree of freedom $\hat{q}_{2}$. Then, for the postselection of the system state $\bra{a_{2l}}$ (which corresponds to one of the eigenstates of the system variable $\hat{A}_{2}$ that is von Neumann coupled with $\hat{q}_{2}$). 
Going through the similar calculations leading up to Eq. (\ref{ss2}) we can obtain the state after the strong von Neumann interaction but before the postselection as follows
\begin{align}
\bra{\psi}& = \Sigma_{k}\int(1+i\lambda_{1}(A_{1})_{w}\hat{q}_{1})\phi_{i}(p_{1},p_{2}-a_{2k})\nonumber\\
\label{ss6}
& \bra{a_{2k}}\bra{p_{1}}\bra{p_{2}}d\vec{p}
\end{align}
Postselecting $\bra{a_{2l}}$ and writing Eq. (\ref{ss6}) in terms of weak values we obtain the postselected pointer state
\begin{equation}
\label{ss7}
\bra{\phi_{f}}\approx \braket{a_{2l}}{\psi_{i}}\int(1+i\lambda_{1}(A_{1})_{w}\hat{q}_{1})\phi_{i}(p_{1},p_{2}-a_{2l}) \bra{p_{1}}\bra{p_{2}}d\vec{p}
\end{equation} 
going through the similar calculations as before, one can obtain the following results for the shifts of the expectation values of the pointer degrees of freedom 
  \begin{equation}
  \label{sqm23}
 \langle \hat{q}_{1} \rangle_{f} = \langle \hat{q}_{1} \rangle_{i}-2\lambda b var(q_{1})_{i}
 \end{equation}
\begin{equation}
\label{sqm24}
 \langle \hat{q}_{2} \rangle_{f} = \langle \hat{q}_{2} \rangle_{i}-2\lambda b corr(q_{1},q_{2})_{i}
 \end{equation}
  \begin{equation}
  \label{sqm25}
 \langle \hat{p}_{1} \rangle_{f} = \langle \hat{p}_{1} \rangle_{i} + \lambda a + m\lambda b\pd{var(q_{1})_{i}}{t}
 \end{equation}
 \begin{equation}
 \label{sqm26}
 \langle \hat{p}_{2} \rangle_{f} = a_{2l}+\langle \hat{p}_{2} \rangle_{i}+2 \lambda b corr(q_{1},p_{2})_{i}
 \end{equation}
 where a and b are respectively the real and the imaginary parts of the weak value of the observable $\hat{A}$ in question.\\
 
 Here also, the effect of entanglement between $q_{1}$ and system observable $A_{1}$ due to weak interaction is manifested through non-zero correlation between the degree of freedom used for postselection and that used in weak interaction - this can be seen from Eqs. (\ref{sqm24}) and (\ref{sqm26})\\
 
 Note that, Eqs. (\ref{sqm23}) and (\ref{sqm25}) do not contain any effect of pointer state correlation and are the same as Jozsa's results given by (\ref{sqm2}) and \ref{sqm3}).On the other hand, it can be seen from Eqs. (\ref{sqm24}) and (\ref{sqm26}) that the effect of pointer state correlation is manifested in the shift of that pointer degree of freedom (say, position) which is involved in the von Neumann coupling used for the projective measurement resulting in postselection, as well as in the shift of its conjugate variable (momentum). Then, Eqs. (\ref{sqm24}) and (\ref{sqm26}) constitute the key extension of Jozsa's results in the case of WM scenario involving single weak interaction that arises essentially from the pointer state correlations.Note that, in these Eqs. (\ref{sqm16}) - (\ref{sqm18.5}), (\ref{sqm19}) - (\ref{sqm21}),(\ref{sqm24}) and (\ref{sqm26}) the correlation terms essentially contain the imaginary part of the weak value. This means that for the effect of the pointer state correlation to be manifested in terms of observable shifts of the pointer degrees of freedom, the relevant weak value has to be necessarily complex.   \\
 
Here it is relevant to mention that in the treatment of weak measurement using orbital angular momentum(OAM) pointer states in terms of the Laguerre-Gaussian optical modes, it has been noted that for the optical modes endowed with OAM, the pointer state distribution is not factorisable \cite{JQV1}. In this context, for extracting the joint weak values from the two-dimensional spatial displacements, the relevant results for the shifts of the mean pointer position degrees of freedom have been obtained by Kobayashi et al. \cite{jqv}.

To put it more specifically, in the aforementioned paper, the preselected pointer state is represented by the two dimensional Laguerre Gauss modes with non-vanishing OAM $l$, and the weak interaction is taken to be of the form 
\begin{equation}
\label{lg1}
H=g\delta(t-t_{0})(\hat{A}\otimes\hat{p_{x}}+\hat{B}\otimes\hat{p_{y}})
\end{equation}
where $g$ is the coupling parameter. The preselected pointer state is given by
\begin{equation}
\label{lg0}
\psi(x,y,l)=N(x+isgn(l)y)^{|l|}e^{-(x^{2}+y^{2})/4\sigma^{2}}
\end{equation}
where $N$ is the normalization constant and $2\sigma$ is the beam waist. Kobayashi et. al. \cite{jqv} considered the weak measurement scheme using the Hamiltonian given by Eq. (\ref{lg1}) for the Laguerre Gauss mode with OAM $l$ as the initial pointer state. They obtained the following relations
  \begin{equation}
  \label{lg4}
  \langle \hat{x} \rangle_{f}-\langle \hat{x} \rangle_{i}=g[Re(A)_{w}+lIm(B)_{w}]
  \end{equation}
  \begin{equation}
  \label{lg5}
  \langle \hat{y} \rangle_{f}-\langle \hat{y} \rangle_{i}=g[Re(B)_{w}-lIm(A)_{w}]
  \end{equation}
   Now, note that the compatibility between Eqs. (\ref{sqm19}),(\ref{sqm20}) (obtained in our more general treatment) and Eqs. (\ref{lg4}), (\ref{lg5}) requires that for the Laguerre Gauss mode with OAM $l$ characterized by $\psi(x,y,l)$ given by Eq. (\ref{lg0})
   \begin{equation}
   \label{lg6}
   corr(p_{x},y)=corr(p_{y},x)=\frac{l}{2}
   \end{equation}
   
where $l$ is the OAM corresponding to the Laguerre Gauss mode. In order to check whether this condition is indeed satisfied, we've calculated the values of correlations for the OAM $l=1$ Laguerre Gauss mode, $\psi(x,y,l=1)$ given by Eq. \ref{lg0}. It is found that for the two dimensional spatial wave function corresponding to the OAM $l=1$ Laguerre Gauss mode , using weak interaction of the form given by Eq. (\ref{lg1}), one obtains $corr(x,y)=0$ and $corr(p_{x},y)=corr(p_{y},x)=\frac{1}{2}$, thereby ensuring in this case the compatibility between Eqs. (\ref{lg4}), (\ref{lg5}) and Eqs. (\ref{sqm19}) and (\ref{sqm20}). Similarly, it can be checked that this compatibility holds for any other value of $l$. Turek et. al.\cite{turek} later generalized the treatment using all orders of interaction strength and by taking specific system observables. If our treatment is extended to include higher orders of interaction strength. It would give rise to higher order cross moments pertaining to pointer degrees of freedom. This calls for further study.\\
 
Here we may stress that we have considered the \textit{generic} case of non-separable pointer state, but in the treatment by Kobayashi et al. \cite{jqv} they consider a specific case of the non-separability of the pointer state arising from non-vanishing orbital angular momentum in the optical modes. A curious point to be noted is that the observable shifts obtained in our treatment of sequential WM by considering each of the two weak interactions to be involving single von Neumann coupling turn out to be the same as that obtained by Kobayashi et al. \cite{jqv} using a single weak interaction with two von Neumann couplings given by Eq. (\ref{lg1}). Since both these treatments are based on considering effects up to the first order of weak interaction, it should be worth investigating in detail the implication of this equivalence and how this is affected by considering effects up to the second order of weak interactions. A further implication of extending our treatment of the sequential WM for the second order of weak interactions would be to probe the way the results obtained in the presence of pointer state correlations would reduce to the results that were derived by Mitchison et al. \citep{SQM} for the second order of weak interactions but without considering the pointer state correlations.\\

\section{Concluding Remarks and Outlook} 
To put it in a nutshell, in the case of sequential WM as well as for the usually considered WM scenario involving single weak interaction, our treatment shows the way the effect of the pointer state correlation is reflected in the empirical manifestation of weak values in terms of observable shifts of the expectation values of the pointer degrees of freedom. In both these cases we have derived explicit forms of the pertinent extensions (Eqs. (\ref{sqm16}) - (\ref{sqm18.5}), (\ref{sqm19}) - (\ref{sqm21}),(\ref{sqm24}) and (\ref{sqm26}))  of Jozsa's original results. A key point to be noted here is that in each of these equations, the effect of correlations among the pointer degrees of freedom is embodied in \textit{those} individual terms which essentially contain the imaginary part of the weak values involved. Thus, for real weak values, the observable shifts of the expectation values of the pointer degrees of freedom do \textit{not} contain any effect of the pointer state correlations.\\

 Now, regarding the role of entanglement between the pointer degrees of freedom in the pointer state $\bra{\phi_{in}}$, we may stress that the existence of non-zero correlation between the pointer degrees of freedom does not necessarily imply entanglement. It is only for the two mode Gaussian wave function, the non-zero correlation between the pointer degrees of freedom necessarily implies entanglement in the sense of satisfying the PPT criterion for continuous variable entanglement \cite{raja}.  For the three mode Gaussian wave function, non-zero correlations do not necessarily imply entanglement \cite{gauss}. Thus, our treatment of the standard WM scenario (Sec. III) involving single weak interaction and the choice of $\bra{\phi_{i}}$ (Eq.(\ref{sqm22})) to be two mode Gaussian involving non-zero correlation has an interesting testable implication in the sense that the shift of the final expectation values of the pointer degrees of freedom will contain the effect of entanglement in terms of the relevant correlations (Eqs. (\ref{sqm24}) and (\ref{sqm26}) of Sec. III). To elaborate on this, note that, a general two-mode Gaussian state can be written as follows
 \begin{equation}
 \label{eq2}
 \psi(q_{1},q_{2})=N exp[-(\alpha q_{1}^{2}+\beta q_{2}^{2}+2\gamma q_{1}q_{2})]
 \end{equation}
  where $N$ is the normalization constant and the above state implies correlation given by $corr(q_{1},q_{2})\propto\gamma$. Using the criterion for any Gaussian state to be entangled \cite{simon}, the value of the determinant of the matrix $C$ defined as follows for the two mode Gaussian determines whether it is entangled 
  \begin{equation}
  \label{eq3}
      C=
     \left[ {\begin{array}{cc}
      <q_{1}q_{2}> & <q_{1}p_{2}> \\       <p_{1}q_{2}> & <p_{1}p_{2}> \      \end{array} } \right]
   \end{equation}
  When $det(C) < 0$, the state is entangled. Using this criterion,
 it has been shown that for all non-zero values of $\gamma$, the two mode Gaussian state given by the Eq.(\ref{eq2}) is entangled \cite{raja}.\\
  
   An important point to note here is that, using Eqs. (\ref{sqm24}) and (\ref{sqm26}) giving the shifts of the expectation values of the postselected pointer degrees of freedom ($q_{2}$ and $p_{2}$), one can obtain the diagonal terms of the matrix $C$ from the experimentally determined shifts. Similarly, in order to empirically determine the off diagonal terms of the matrix $C$, one needs to interchange the pointer degrees of freedom involved in weak interaction and projective measurement used in the postselection. Thus, by determining all the elements of the matrix $C$, one can verify whether the initial pointer state in question is entangled or not. Note that this procedure essentially uses the scheme developed in Sec.III of our paper.\\
   
    Similarly, for the three-mode Gaussian as the initial pointer state, one can obtain the elements of the matrix $C$ using the scheme presented in Sec.II for the sequential WM. An extension of this procedure seems possible for the initial pointer state taken to be multi-mode Gaussian. A comprehensive investigation of this possibility will be pursued in a sequel paper. \\
 
  Another line of future study could be with respect to the paper mentioned earlier by G. Mitchison \cite{cum}. In that paper, joint expectation values of the  postselected pointer degrees of freedom are obtained in terms of joint weak values for factorizable multidimensional preselected pointer state, i.e., in the absence of the correlation between different pointer degrees of freedom. One can thus investigate how their result would be modified if we consider the initial preselected pointer state to be correlated among different pointer degrees of freedom.\\
     
     It may also be instructive to probe how correlations in the initial pointer state would affect the earlier analysis of "weak trajectories" extracted from the pointers of a series of weakly interacting devices using factorizable N- dimensional Gaussian function for the initial pointer state \cite{alexprl}.
  Finally, we offer a few remarks about a possible experimental test of the effect of pointer state correlations derived in the present paper. Recently, predictions made by Mitchison et al. \cite{SQM} have been confirmed by Piacentini et al. \cite{marco} in an interesting experiment by considering sequential weak interactions undergone by single photons using birefringence in optical crystals. In this setup, the preselected two dimensional pointer state is experimentally prepared to be a separable Gaussian, by using single photon guided in a single-mode optical fiber that is suitably collimated with a telescopic optical system. It would therefore be an interesting extension of this experimental setup to  study how the shifts of the postselected probability distribution functions of the relevant pointer variables are modified by an appropriate tuning of the procedure for preparing the preselected pointer state that would entail pointer state correlations. Such an investigation would thus constitute a critical empirical check of the results that have been obtained and analyzed in this paper.

\section{Acknowledgements}
We thank Lev Vaidman for helpful comments on the initial version of this paper. Thanks are also due to the anonymous referee for insightful suggestions. the visit of GM enabling this collaboration was supported by the DST INSPIRE Fellowship. The research of DH is supported by the Dept. of Science and Technology (DST), Govt. of India and Center for Science, Kolkata. SK acknowledges the support of Fellowship of Bose Institute, Kolkata.

\bibliography{13thnov15}  

\begin{thebibliography}{49}%
\makeatletter
\providecommand \@ifxundefined [1]{%
 \@ifx{#1\undefined}
}%
\providecommand \@ifnum [1]{%
 \ifnum #1\expandafter \@firstoftwo
 \else \expandafter \@secondoftwo
 \fi
}%
\providecommand \@ifx [1]{%
 \ifx #1\expandafter \@firstoftwo
 \else \expandafter \@secondoftwo
 \fi
}%
\providecommand \natexlab [1]{#1}%
\providecommand \enquote  [1]{``#1''}%
\providecommand \bibnamefont  [1]{#1}%
\providecommand \bibfnamefont [1]{#1}%
\providecommand \citenamefont [1]{#1}%
\providecommand \href@noop [0]{\@secondoftwo}%
\providecommand \href [0]{\begingroup \@sanitize@url \@href}%
\providecommand \@href[1]{\@@startlink{#1}\@@href}%
\providecommand \@@href[1]{\endgroup#1\@@endlink}%
\providecommand \@sanitize@url [0]{\catcode `\\12\catcode `\$12\catcode
  `\&12\catcode `\#12\catcode `\^12\catcode `\_12\catcode `\%12\relax}%
\providecommand \@@startlink[1]{}%
\providecommand \@@endlink[0]{}%
\providecommand \url  [0]{\begingroup\@sanitize@url \@url }%
\providecommand \@url [1]{\endgroup\@href {#1}{\urlprefix }}%
\providecommand \urlprefix  [0]{URL }%
\providecommand \Eprint [0]{\href }%
\providecommand \doibase [0]{http://dx.doi.org/}%
\providecommand \selectlanguage [0]{\@gobble}%
\providecommand \bibinfo  [0]{\@secondoftwo}%
\providecommand \bibfield  [0]{\@secondoftwo}%
\providecommand \translation [1]{[#1]}%
\providecommand \BibitemOpen [0]{}%
\providecommand \bibitemStop [0]{}%
\providecommand \bibitemNoStop [0]{.\EOS\space}%
\providecommand \EOS [0]{\spacefactor3000\relax}%
\providecommand \BibitemShut  [1]{\csname bibitem#1\endcsname}%
\let\auto@bib@innerbib\@empty
\bibitem [{\citenamefont {Aharonov}\ \emph {et~al.}(1988)\citenamefont
  {Aharonov}, \citenamefont {Albert},\ and\ \citenamefont
  {Vaidman}}]{aharonov}%
  \BibitemOpen
  \bibfield  {author} {\bibinfo {author} {\bibfnamefont {Y.}~\bibnamefont
  {Aharonov}}, \bibinfo {author} {\bibfnamefont {D.~Z.}\ \bibnamefont
  {Albert}}, \ and\ \bibinfo {author} {\bibfnamefont {L.}~\bibnamefont
  {Vaidman}},\ }\href {\doibase 10.1103/PhysRevLett.60.1351} {\bibfield
  {journal} {\bibinfo  {journal} {Phys. Rev. Lett.}\ }\textbf {\bibinfo
  {volume} {60}},\ \bibinfo {pages} {1351} (\bibinfo {year}
  {1988})}\BibitemShut {NoStop}%
\bibitem [{\citenamefont {Leggett}(1989)}]{legget}%
  \BibitemOpen
  \bibfield  {author} {\bibinfo {author} {\bibfnamefont {A.~J.}\ \bibnamefont
  {Leggett}},\ }\href {\doibase 10.1103/PhysRevLett.62.2325} {\bibfield
  {journal} {\bibinfo  {journal} {Phys. Rev. Lett.}\ }\textbf {\bibinfo
  {volume} {62}},\ \bibinfo {pages} {2325} (\bibinfo {year}
  {1989})}\BibitemShut {NoStop}%
\bibitem [{\citenamefont {Peres}(1989)}]{peres}%
  \BibitemOpen
  \bibfield  {author} {\bibinfo {author} {\bibfnamefont {A.}~\bibnamefont
  {Peres}},\ }\href@noop {} {\bibfield  {journal} {\bibinfo  {journal} {Phys.
  Rev. Lett.}\ }\textbf {\bibinfo {volume} {62}},\ \bibinfo {pages} {2326}
  (\bibinfo {year} {1989})}\BibitemShut {NoStop}%
\bibitem [{\citenamefont {Aharonov}\ and\ \citenamefont
  {Vaidman}(1989)}]{aharonov1}%
  \BibitemOpen
  \bibfield  {author} {\bibinfo {author} {\bibfnamefont {Y.}~\bibnamefont
  {Aharonov}}\ and\ \bibinfo {author} {\bibfnamefont {L.}~\bibnamefont
  {Vaidman}},\ }\href@noop {} {\bibfield  {journal} {\bibinfo  {journal} {Phys.
  Rev. Lett.}\ }\textbf {\bibinfo {volume} {62}},\ \bibinfo {pages} {2327}
  (\bibinfo {year} {1989})}\BibitemShut {NoStop}%
\bibitem [{\citenamefont {Duck}\ \emph {et~al.}(1989)\citenamefont {Duck},
  \citenamefont {Stevenson},\ and\ \citenamefont {Sudarshan}}]{Duck}%
  \BibitemOpen
  \bibfield  {author} {\bibinfo {author} {\bibfnamefont {I.~M.}\ \bibnamefont
  {Duck}}, \bibinfo {author} {\bibfnamefont {P.~M.}\ \bibnamefont {Stevenson}},
  \ and\ \bibinfo {author} {\bibfnamefont {E.~C.~G.}\ \bibnamefont
  {Sudarshan}},\ }\href {\doibase 10.1103/PhysRevD.40.2112} {\bibfield
  {journal} {\bibinfo  {journal} {Phys. Rev. D}\ }\textbf {\bibinfo {volume}
  {40}},\ \bibinfo {pages} {2112} (\bibinfo {year} {1989})}\BibitemShut
  {NoStop}%
\bibitem [{\citenamefont {Aharonov}\ and\ \citenamefont
  {Vaidman}(1991)}]{aharonov2}%
  \BibitemOpen
  \bibfield  {author} {\bibinfo {author} {\bibfnamefont {Y.}~\bibnamefont
  {Aharonov}}\ and\ \bibinfo {author} {\bibfnamefont {L.}~\bibnamefont
  {Vaidman}},\ }\href@noop {} {\bibfield  {journal} {\bibinfo  {journal} {J.
  Phys. A}\ }\textbf {\bibinfo {volume} {24}},\ \bibinfo {pages} {2315}
  (\bibinfo {year} {1991})}\BibitemShut {NoStop}%
\bibitem [{\citenamefont {Aharonov}\ and\ \citenamefont
  {Vaidman}(1990)}]{vaid}%
  \BibitemOpen
  \bibfield  {author} {\bibinfo {author} {\bibfnamefont {Y.}~\bibnamefont
  {Aharonov}}\ and\ \bibinfo {author} {\bibfnamefont {L.}~\bibnamefont
  {Vaidman}},\ }\href {\doibase 10.1103/PhysRevA.41.11} {\bibfield  {journal}
  {\bibinfo  {journal} {Phys. Rev. A}\ }\textbf {\bibinfo {volume} {41}},\
  \bibinfo {pages} {11} (\bibinfo {year} {1990})}\BibitemShut {NoStop}%
\bibitem [{\citenamefont {Tollaksen}(2007)}]{tollak}%
  \BibitemOpen
  \bibfield  {author} {\bibinfo {author} {\bibfnamefont {J.}~\bibnamefont
  {Tollaksen}},\ }\href@noop {} {\bibfield  {journal} {\bibinfo  {journal} {J.
  Phys. A}\ }\textbf {\bibinfo {volume} {40}},\ \bibinfo {pages} {9033}
  (\bibinfo {year} {2007})}\BibitemShut {NoStop}%
\bibitem [{\citenamefont {Brunner}\ and\ \citenamefont
  {Simon}(2010)}]{brunner}%
  \BibitemOpen
  \bibfield  {author} {\bibinfo {author} {\bibfnamefont {N.}~\bibnamefont
  {Brunner}}\ and\ \bibinfo {author} {\bibfnamefont {C.}~\bibnamefont
  {Simon}},\ }\href@noop {} {\bibfield  {journal} {\bibinfo  {journal} {Phys.
  Rev. Lett.}\ }\textbf {\bibinfo {volume} {105}},\ \bibinfo {pages} {010405}
  (\bibinfo {year} {2010})}\BibitemShut {NoStop}%
\bibitem [{\citenamefont {Pan}\ and\ \citenamefont {Matzkin}(2012)}]{WSG}%
  \BibitemOpen
  \bibfield  {author} {\bibinfo {author} {\bibfnamefont {A.~K.}\ \bibnamefont
  {Pan}}\ and\ \bibinfo {author} {\bibfnamefont {A.}~\bibnamefont {Matzkin}},\
  }\href {\doibase 10.1103/PhysRevA.85.022122} {\bibfield  {journal} {\bibinfo
  {journal} {Phys. Rev. A}\ }\textbf {\bibinfo {volume} {85}},\ \bibinfo
  {pages} {022122} (\bibinfo {year} {2012})}\BibitemShut {NoStop}%
\bibitem [{\citenamefont {{Hari Dass}}()}]{haridass}%
  \BibitemOpen
  \bibfield  {author} {\bibinfo {author} {\bibfnamefont {N.~D.}\ \bibnamefont
  {{Hari Dass}}},\ }\href@noop {} {\ }\Eprint {http://arxiv.org/abs/1509.04869}
  {arXiv:1509.04869 [quant-ph]} \BibitemShut {NoStop}%
\bibitem [{\citenamefont {Dressel}\ \emph {et~al.}(2014)\citenamefont
  {Dressel}, \citenamefont {Malik}, \citenamefont {Miatto}, \citenamefont
  {Jordan},\ and\ \citenamefont {Boyd}}]{dress}%
  \BibitemOpen
  \bibfield  {author} {\bibinfo {author} {\bibfnamefont {J.}~\bibnamefont
  {Dressel}}, \bibinfo {author} {\bibfnamefont {M.}~\bibnamefont {Malik}},
  \bibinfo {author} {\bibfnamefont {F.~M.}\ \bibnamefont {Miatto}}, \bibinfo
  {author} {\bibfnamefont {A.~N.}\ \bibnamefont {Jordan}}, \ and\ \bibinfo
  {author} {\bibfnamefont {W.}~\bibnamefont {Boyd}, \bibfnamefont {Robert}},\
  }\href {\doibase 10.1103/RevModPhys.86.307} {\bibfield  {journal} {\bibinfo
  {journal} {Rev. Mod. Phys.}\ }\textbf {\bibinfo {volume} {86}},\ \bibinfo
  {pages} {307} (\bibinfo {year} {2014})}\BibitemShut {NoStop}%
\bibitem [{\citenamefont {Yokota}\ \emph {et~al.}(2009)\citenamefont {Yokota},
  \citenamefont {Yamamoto}, \citenamefont {Koashi},\ and\ \citenamefont
  {Imoto}}]{hardy}%
  \BibitemOpen
  \bibfield  {author} {\bibinfo {author} {\bibfnamefont {K.}~\bibnamefont
  {Yokota}}, \bibinfo {author} {\bibfnamefont {T.}~\bibnamefont {Yamamoto}},
  \bibinfo {author} {\bibfnamefont {M.}~\bibnamefont {Koashi}}, \ and\ \bibinfo
  {author} {\bibfnamefont {N.}~\bibnamefont {Imoto}},\ }\href
  {http://stacks.iop.org/1367-2630/11/i=3/a=033011} {\bibfield  {journal}
  {\bibinfo  {journal} {New J Phys}\ }\textbf {\bibinfo {volume} {11}},\
  \bibinfo {pages} {033011} (\bibinfo {year} {2009})}\BibitemShut {NoStop}%
\bibitem [{\citenamefont {Matzkin}\ and\ \citenamefont {Pan}(2013)}]{3-box}%
  \BibitemOpen
  \bibfield  {author} {\bibinfo {author} {\bibfnamefont {A.}~\bibnamefont
  {Matzkin}}\ and\ \bibinfo {author} {\bibfnamefont {A.~K.}\ \bibnamefont
  {Pan}},\ }\href {http://stacks.iop.org/1751-8121/46/i=31/a=315307} {\bibfield
   {journal} {\bibinfo  {journal} {J. Phy. A: Math. Theor.}\ }\textbf {\bibinfo
  {volume} {46}},\ \bibinfo {pages} {315307} (\bibinfo {year}
  {2013})}\BibitemShut {NoStop}%
\bibitem [{\citenamefont {Aharonov}\ \emph {et~al.}(2013)\citenamefont
  {Aharonov}, \citenamefont {Popescu}, \citenamefont {Rohrlich},\ and\
  \citenamefont {Skrzypczyk}}]{QCC}%
  \BibitemOpen
  \bibfield  {author} {\bibinfo {author} {\bibfnamefont {Y.}~\bibnamefont
  {Aharonov}}, \bibinfo {author} {\bibfnamefont {S.}~\bibnamefont {Popescu}},
  \bibinfo {author} {\bibfnamefont {D.}~\bibnamefont {Rohrlich}}, \ and\
  \bibinfo {author} {\bibfnamefont {P.}~\bibnamefont {Skrzypczyk}},\ }\href
  {http://stacks.iop.org/1367-2630/15/i=11/a=113015} {\bibfield  {journal}
  {\bibinfo  {journal} {New J Phys}\ }\textbf {\bibinfo {volume} {15}},\
  \bibinfo {pages} {113015} (\bibinfo {year} {2013})}\BibitemShut {NoStop}%
\bibitem [{\citenamefont {Agarwal}\ and\ \citenamefont {Pathak}(2007)}]{ent}%
  \BibitemOpen
  \bibfield  {author} {\bibinfo {author} {\bibfnamefont {G.~S.}\ \bibnamefont
  {Agarwal}}\ and\ \bibinfo {author} {\bibfnamefont {P.~K.}\ \bibnamefont
  {Pathak}},\ }\href {\doibase http://dx.doi.org/10.1103/PhysRevA.75.032108}
  {\bibfield  {journal} {\bibinfo  {journal} {Phys. Rev. A}\ }\textbf {\bibinfo
  {volume} {75}},\ \bibinfo {pages} {032108} (\bibinfo {year}
  {2007})}\BibitemShut {NoStop}%
\bibitem [{\citenamefont {Kaneda}\ \emph {et~al.}(2014)\citenamefont {Kaneda},
  \citenamefont {Baek}, \citenamefont {Ozawa},\ and\ \citenamefont
  {Edamatsu}}]{kane}%
  \BibitemOpen
  \bibfield  {author} {\bibinfo {author} {\bibfnamefont {F.}~\bibnamefont
  {Kaneda}}, \bibinfo {author} {\bibfnamefont {S.-Y.}\ \bibnamefont {Baek}},
  \bibinfo {author} {\bibfnamefont {M.}~\bibnamefont {Ozawa}}, \ and\ \bibinfo
  {author} {\bibfnamefont {K.}~\bibnamefont {Edamatsu}},\ }\href {\doibase
  10.1103/PhysRevLett.112.020402} {\bibfield  {journal} {\bibinfo  {journal}
  {Phys. Rev. Lett.}\ }\textbf {\bibinfo {volume} {112}},\ \bibinfo {pages}
  {020402} (\bibinfo {year} {2014})}\BibitemShut {NoStop}%
\bibitem [{\citenamefont {Matzkin}(2012)}]{alexprl}%
  \BibitemOpen
  \bibfield  {author} {\bibinfo {author} {\bibfnamefont {A.}~\bibnamefont
  {Matzkin}},\ }\href {\doibase 10.1103/PhysRevLett.109.150407} {\bibfield
  {journal} {\bibinfo  {journal} {Phys. Rev. Lett.}\ }\textbf {\bibinfo
  {volume} {109}},\ \bibinfo {pages} {150407} (\bibinfo {year}
  {2012})}\BibitemShut {NoStop}%
\bibitem [{\citenamefont {Dressel}\ and\ \citenamefont
  {Korotkov}(2014)}]{bell1}%
  \BibitemOpen
  \bibfield  {author} {\bibinfo {author} {\bibfnamefont {J.}~\bibnamefont
  {Dressel}}\ and\ \bibinfo {author} {\bibfnamefont {A.~N.}\ \bibnamefont
  {Korotkov}},\ }\href {\doibase 10.1103/PhysRevA.89.012125} {\bibfield
  {journal} {\bibinfo  {journal} {Phys. Rev. A}\ }\textbf {\bibinfo {volume}
  {89}},\ \bibinfo {pages} {012125} (\bibinfo {year} {2014})}\BibitemShut
  {NoStop}%
\bibitem [{\citenamefont {Palacios-Laloy}\ \emph {et~al.}(2010)\citenamefont
  {Palacios-Laloy}, \citenamefont {Mallet}, \citenamefont {Nguyen},
  \citenamefont {Bertet}, \citenamefont {Vion}, \citenamefont {Esteve},\ and\
  \citenamefont {Korotkov}}]{bell2}%
  \BibitemOpen
  \bibfield  {author} {\bibinfo {author} {\bibfnamefont {A.}~\bibnamefont
  {Palacios-Laloy}}, \bibinfo {author} {\bibfnamefont {F.}~\bibnamefont
  {Mallet}}, \bibinfo {author} {\bibfnamefont {F.}~\bibnamefont {Nguyen}},
  \bibinfo {author} {\bibfnamefont {P.}~\bibnamefont {Bertet}}, \bibinfo
  {author} {\bibfnamefont {D.}~\bibnamefont {Vion}}, \bibinfo {author}
  {\bibfnamefont {D.}~\bibnamefont {Esteve}}, \ and\ \bibinfo {author}
  {\bibfnamefont {A.}~\bibnamefont {Korotkov}},\ }\href {\doibase
  10.1038/nphys1641} {\bibfield  {journal} {\bibinfo  {journal} {Nat. Phys.}\
  }\textbf {\bibinfo {volume} {6}},\ \bibinfo {pages} {442–447} (\bibinfo
  {year} {2010})}\BibitemShut {NoStop}%
\bibitem [{\citenamefont {Singh}\ and\ \citenamefont {Pati}(2014)}]{pati}%
  \BibitemOpen
  \bibfield  {author} {\bibinfo {author} {\bibfnamefont {U.}~\bibnamefont
  {Singh}}\ and\ \bibinfo {author} {\bibfnamefont {A.~K.}\ \bibnamefont
  {Pati}},\ }\href {\doibase http://dx.doi.org/10.1016/j.aop.2014.02.004}
  {\bibfield  {journal} {\bibinfo  {journal} {Ann Phys-NY}\ }\textbf {\bibinfo
  {volume} {343}},\ \bibinfo {pages} {141 } (\bibinfo {year}
  {2014})}\BibitemShut {NoStop}%
\bibitem [{\citenamefont {Pusey}(2014)}]{pusey}%
  \BibitemOpen
  \bibfield  {author} {\bibinfo {author} {\bibfnamefont {M.~F.}\ \bibnamefont
  {Pusey}},\ }\href {\doibase 10.1103/PhysRevLett.113.200401} {\bibfield
  {journal} {\bibinfo  {journal} {Phys. Rev. Lett.}\ }\textbf {\bibinfo
  {volume} {113}},\ \bibinfo {pages} {200401} (\bibinfo {year}
  {2014})}\BibitemShut {NoStop}%
\bibitem [{\citenamefont {Steinberg}(1995)}]{stein1}%
  \BibitemOpen
  \bibfield  {author} {\bibinfo {author} {\bibfnamefont {A.~M.}\ \bibnamefont
  {Steinberg}},\ }\href {\doibase 10.1111/j.1749-6632.1995.tb39044.x}
  {\bibfield  {journal} {\bibinfo  {journal} {Ann NY Acad Sci}\ }\textbf
  {\bibinfo {volume} {755}},\ \bibinfo {pages} {900} (\bibinfo {year}
  {1995})}\BibitemShut {NoStop}%
\bibitem [{\citenamefont {Ahnert}\ and\ \citenamefont {Payne}(2004)}]{arrival}%
  \BibitemOpen
  \bibfield  {author} {\bibinfo {author} {\bibfnamefont {S.~E.}\ \bibnamefont
  {Ahnert}}\ and\ \bibinfo {author} {\bibfnamefont {M.~C.}\ \bibnamefont
  {Payne}},\ }\href {\doibase 10.1103/PhysRevA.69.042103} {\bibfield  {journal}
  {\bibinfo  {journal} {Phys. Rev. A}\ }\textbf {\bibinfo {volume} {69}},\
  \bibinfo {pages} {042103} (\bibinfo {year} {2004})}\BibitemShut {NoStop}%
\bibitem [{\citenamefont {Brunner}\ \emph {et~al.}(2003)\citenamefont
  {Brunner}, \citenamefont {Ac\'{i}n}, \citenamefont {Collins}, \citenamefont
  {Gisin},\ and\ \citenamefont {Scarani}}]{tele}%
  \BibitemOpen
  \bibfield  {author} {\bibinfo {author} {\bibfnamefont {N.}~\bibnamefont
  {Brunner}}, \bibinfo {author} {\bibfnamefont {A.}~\bibnamefont {Ac\'{i}n}},
  \bibinfo {author} {\bibfnamefont {D.}~\bibnamefont {Collins}}, \bibinfo
  {author} {\bibfnamefont {N.}~\bibnamefont {Gisin}}, \ and\ \bibinfo {author}
  {\bibfnamefont {V.}~\bibnamefont {Scarani}},\ }\href {\doibase
  10.1103/PhysRevLett.91.180402} {\bibfield  {journal} {\bibinfo  {journal}
  {Phys. Rev. Lett.}\ }\textbf {\bibinfo {volume} {91}},\ \bibinfo {pages}
  {180402} (\bibinfo {year} {2003})}\BibitemShut {NoStop}%
\bibitem [{\citenamefont {Sponar}\ \emph {et~al.}(2015)\citenamefont {Sponar},
  \citenamefont {Denkmayr}, \citenamefont {Geppert}, \citenamefont {Lemmel},
  \citenamefont {Matzkin}, \citenamefont {Tollaksen},\ and\ \citenamefont
  {Hasegawa}}]{yuji}%
  \BibitemOpen
  \bibfield  {author} {\bibinfo {author} {\bibfnamefont {S.}~\bibnamefont
  {Sponar}}, \bibinfo {author} {\bibfnamefont {T.}~\bibnamefont {Denkmayr}},
  \bibinfo {author} {\bibfnamefont {H.}~\bibnamefont {Geppert}}, \bibinfo
  {author} {\bibfnamefont {H.}~\bibnamefont {Lemmel}}, \bibinfo {author}
  {\bibfnamefont {A.}~\bibnamefont {Matzkin}}, \bibinfo {author} {\bibfnamefont
  {J.}~\bibnamefont {Tollaksen}}, \ and\ \bibinfo {author} {\bibfnamefont
  {Y.}~\bibnamefont {Hasegawa}},\ }\href {\doibase 10.1103/PhysRevA.92.062121}
  {\bibfield  {journal} {\bibinfo  {journal} {Phys. Rev. A}\ }\textbf {\bibinfo
  {volume} {92}},\ \bibinfo {pages} {062121} (\bibinfo {year}
  {2015})}\BibitemShut {NoStop}%
\bibitem [{\citenamefont {Zhang}\ \emph {et~al.}(2015)\citenamefont {Zhang},
  \citenamefont {Datta},\ and\ \citenamefont {Walmsley}}]{metro}%
  \BibitemOpen
  \bibfield  {author} {\bibinfo {author} {\bibfnamefont {L.}~\bibnamefont
  {Zhang}}, \bibinfo {author} {\bibfnamefont {A.}~\bibnamefont {Datta}}, \ and\
  \bibinfo {author} {\bibfnamefont {I.~A.}\ \bibnamefont {Walmsley}},\ }\href
  {\doibase 10.1103/PhysRevLett.114.210801} {\bibfield  {journal} {\bibinfo
  {journal} {Phys. Rev. Lett.}\ }\textbf {\bibinfo {volume} {114}},\ \bibinfo
  {pages} {210801} (\bibinfo {year} {2015})}\BibitemShut {NoStop}%
\bibitem [{\citenamefont {Hosten}\ and\ \citenamefont
  {Kwiat}(2008)}]{spinhall}%
  \BibitemOpen
  \bibfield  {author} {\bibinfo {author} {\bibfnamefont {O.}~\bibnamefont
  {Hosten}}\ and\ \bibinfo {author} {\bibfnamefont {P.}~\bibnamefont {Kwiat}},\
  }\href {\doibase 10.1126/science.1152697} {\bibfield  {journal} {\bibinfo
  {journal} {Science}\ }\textbf {\bibinfo {volume} {319}},\ \bibinfo {pages}
  {787} (\bibinfo {year} {2008})}\BibitemShut {NoStop}%
\bibitem [{\citenamefont {Starling}\ \emph {et~al.}(2009)\citenamefont
  {Starling}, \citenamefont {Dixon}, \citenamefont {Jordan},\ and\
  \citenamefont {Howell}}]{TBD}%
  \BibitemOpen
  \bibfield  {author} {\bibinfo {author} {\bibfnamefont {D.~J.}\ \bibnamefont
  {Starling}}, \bibinfo {author} {\bibfnamefont {P.~B.}\ \bibnamefont {Dixon}},
  \bibinfo {author} {\bibfnamefont {A.~N.}\ \bibnamefont {Jordan}}, \ and\
  \bibinfo {author} {\bibfnamefont {J.~C.}\ \bibnamefont {Howell}},\ }\href
  {\doibase 10.1103/PhysRevA.80.041803} {\bibfield  {journal} {\bibinfo
  {journal} {Phys. Rev. A}\ }\textbf {\bibinfo {volume} {80}},\ \bibinfo
  {pages} {041803} (\bibinfo {year} {2009})}\BibitemShut {NoStop}%
\bibitem [{\citenamefont {Goswami}\ \emph {et~al.}(2014)\citenamefont
  {Goswami}, \citenamefont {Pal}, \citenamefont {Nandi}, \citenamefont
  {Panigrahi},\ and\ \citenamefont {Ghosh}}]{Goswami:14}%
  \BibitemOpen
  \bibfield  {author} {\bibinfo {author} {\bibfnamefont {S.}~\bibnamefont
  {Goswami}}, \bibinfo {author} {\bibfnamefont {M.}~\bibnamefont {Pal}},
  \bibinfo {author} {\bibfnamefont {A.}~\bibnamefont {Nandi}}, \bibinfo
  {author} {\bibfnamefont {P.~K.}\ \bibnamefont {Panigrahi}}, \ and\ \bibinfo
  {author} {\bibfnamefont {N.}~\bibnamefont {Ghosh}},\ }\href {\doibase
  10.1364/OL.39.006229} {\bibfield  {journal} {\bibinfo  {journal} {Opt.
  Lett.}\ }\textbf {\bibinfo {volume} {39}},\ \bibinfo {pages} {6229} (\bibinfo
  {year} {2014})}\BibitemShut {NoStop}%
\bibitem [{\citenamefont {Salazar-Serrano}\ \emph {et~al.}(2014)\citenamefont
  {Salazar-Serrano}, \citenamefont {Janner}, \citenamefont {Brunner},
  \citenamefont {Pruneri},\ and\ \citenamefont {Torres}}]{TS}%
  \BibitemOpen
  \bibfield  {author} {\bibinfo {author} {\bibfnamefont {L.~J.}\ \bibnamefont
  {Salazar-Serrano}}, \bibinfo {author} {\bibfnamefont {D.}~\bibnamefont
  {Janner}}, \bibinfo {author} {\bibfnamefont {N.}~\bibnamefont {Brunner}},
  \bibinfo {author} {\bibfnamefont {V.}~\bibnamefont {Pruneri}}, \ and\
  \bibinfo {author} {\bibfnamefont {J.~P.}\ \bibnamefont {Torres}},\ }\href
  {\doibase 10.1103/PhysRevA.89.012126} {\bibfield  {journal} {\bibinfo
  {journal} {Phys. Rev. A}\ }\textbf {\bibinfo {volume} {89}},\ \bibinfo
  {pages} {012126} (\bibinfo {year} {2014})}\BibitemShut {NoStop}%
\bibitem [{\citenamefont {Lundeen}\ \emph {et~al.}(2011)\citenamefont
  {Lundeen}, \citenamefont {Sutherland}, \citenamefont {Patel}, \citenamefont
  {Stewart},\ and\ \citenamefont {Bamber}}]{QWF}%
  \BibitemOpen
  \bibfield  {author} {\bibinfo {author} {\bibfnamefont {J.~S.}\ \bibnamefont
  {Lundeen}}, \bibinfo {author} {\bibfnamefont {S.}~\bibnamefont {Sutherland}},
  \bibinfo {author} {\bibfnamefont {A.}~\bibnamefont {Patel}}, \bibinfo
  {author} {\bibfnamefont {C.}~\bibnamefont {Stewart}}, \ and\ \bibinfo
  {author} {\bibfnamefont {C.}~\bibnamefont {Bamber}},\ }\href {\doibase
  10.1038/nature10120} {\bibfield  {journal} {\bibinfo  {journal} {Nature}\
  }\textbf {\bibinfo {volume} {474}},\ \bibinfo {pages} {188–191} (\bibinfo
  {year} {2011})}\BibitemShut {NoStop}%
\bibitem [{\citenamefont {Kocsis}\ \emph {et~al.}(2011)\citenamefont {Kocsis},
  \citenamefont {Braverman}, \citenamefont {Ravets}, \citenamefont {Stevens},
  \citenamefont {Mirin}, \citenamefont {Shalm},\ and\ \citenamefont
  {Steinberg}}]{stein}%
  \BibitemOpen
  \bibfield  {author} {\bibinfo {author} {\bibfnamefont {S.}~\bibnamefont
  {Kocsis}}, \bibinfo {author} {\bibfnamefont {B.}~\bibnamefont {Braverman}},
  \bibinfo {author} {\bibfnamefont {S.}~\bibnamefont {Ravets}}, \bibinfo
  {author} {\bibfnamefont {M.~J.}\ \bibnamefont {Stevens}}, \bibinfo {author}
  {\bibfnamefont {R.~P.}\ \bibnamefont {Mirin}}, \bibinfo {author}
  {\bibfnamefont {L.~K.}\ \bibnamefont {Shalm}}, \ and\ \bibinfo {author}
  {\bibfnamefont {A.~M.}\ \bibnamefont {Steinberg}},\ }\href {\doibase
  10.1126/science.1202218} {\bibfield  {journal} {\bibinfo  {journal}
  {science}\ }\textbf {\bibinfo {volume} {332}},\ \bibinfo {pages} {1170}
  (\bibinfo {year} {2011})}\BibitemShut {NoStop}%
\bibitem [{\citenamefont {Jozsa}(2007)}]{jozsa}%
  \BibitemOpen
  \bibfield  {author} {\bibinfo {author} {\bibfnamefont {R.}~\bibnamefont
  {Jozsa}},\ }\href {\doibase 10.1103/PhysRevA.76.044103} {\bibfield  {journal}
  {\bibinfo  {journal} {Phys. Rev. A}\ }\textbf {\bibinfo {volume} {76}},\
  \bibinfo {pages} {044103} (\bibinfo {year} {2007})}\BibitemShut {NoStop}%
\bibitem [{\citenamefont {Resch}\ and\ \citenamefont
  {Steinberg}(2004)}]{stein3}%
  \BibitemOpen
  \bibfield  {author} {\bibinfo {author} {\bibfnamefont {K.~J.}\ \bibnamefont
  {Resch}}\ and\ \bibinfo {author} {\bibfnamefont {A.~M.}\ \bibnamefont
  {Steinberg}},\ }\href {\doibase 10.1103/PhysRevLett.92.130402} {\bibfield
  {journal} {\bibinfo  {journal} {Phys. Rev. Lett.}\ }\textbf {\bibinfo
  {volume} {92}},\ \bibinfo {pages} {130402} (\bibinfo {year}
  {2004})}\BibitemShut {NoStop}%
\bibitem [{\citenamefont {Resch}(2004)}]{prac}%
  \BibitemOpen
  \bibfield  {author} {\bibinfo {author} {\bibfnamefont {K.~J.}\ \bibnamefont
  {Resch}},\ }\href {http://stacks.iop.org/1464-4266/6/i=11/a=009} {\bibfield
  {journal} {\bibinfo  {journal} {Journal of Optics B: Quantum and
  Semiclassical Optics}\ }\textbf {\bibinfo {volume} {6}},\ \bibinfo {pages}
  {482} (\bibinfo {year} {2004})}\BibitemShut {NoStop}%
\bibitem [{\citenamefont {Mitchison}(2008)}]{cum}%
  \BibitemOpen
  \bibfield  {author} {\bibinfo {author} {\bibfnamefont {G.}~\bibnamefont
  {Mitchison}},\ }\href {\doibase 10.1103/PhysRevA.77.052102} {\bibfield
  {journal} {\bibinfo  {journal} {Phys. Rev. A}\ }\textbf {\bibinfo {volume}
  {77}},\ \bibinfo {pages} {052102} (\bibinfo {year} {2008})}\BibitemShut
  {NoStop}%
\bibitem [{\citenamefont {Lundeen}\ and\ \citenamefont {Resch}(2005)}]{prac1}%
  \BibitemOpen
  \bibfield  {author} {\bibinfo {author} {\bibfnamefont {J.}~\bibnamefont
  {Lundeen}}\ and\ \bibinfo {author} {\bibfnamefont {K.}~\bibnamefont
  {Resch}},\ }\href {\doibase http://dx.doi.org/10.1016/j.physleta.2004.11.037}
  {\bibfield  {journal} {\bibinfo  {journal} {Physics Letters A}\ }\textbf
  {\bibinfo {volume} {334}},\ \bibinfo {pages} {337 } (\bibinfo {year}
  {2005})}\BibitemShut {NoStop}%
\bibitem [{\citenamefont {de~Lima~Bernardo}\ \emph {et~al.}(2014)\citenamefont
  {de~Lima~Bernardo}, \citenamefont {Azevedo},\ and\ \citenamefont
  {Rosas}}]{JQV2}%
  \BibitemOpen
  \bibfield  {author} {\bibinfo {author} {\bibfnamefont {B.}~\bibnamefont
  {de~Lima~Bernardo}}, \bibinfo {author} {\bibfnamefont {S.}~\bibnamefont
  {Azevedo}}, \ and\ \bibinfo {author} {\bibfnamefont {A.}~\bibnamefont
  {Rosas}},\ }\href {\doibase http://dx.doi.org/10.1016/j.optcom.2014.06.008}
  {\bibfield  {journal} {\bibinfo  {journal} {Opt. Commun.}\ }\textbf {\bibinfo
  {volume} {331}},\ \bibinfo {pages} {194 } (\bibinfo {year}
  {2014})}\BibitemShut {NoStop}%
\bibitem [{\citenamefont {Di~Lorenzo}\ and\ \citenamefont
  {Egues}(2008)}]{lorenzo}%
  \BibitemOpen
  \bibfield  {author} {\bibinfo {author} {\bibfnamefont {A.}~\bibnamefont
  {Di~Lorenzo}}\ and\ \bibinfo {author} {\bibfnamefont {J.~C.}\ \bibnamefont
  {Egues}},\ }\href {\doibase 10.1103/PhysRevA.77.042108} {\bibfield  {journal}
  {\bibinfo  {journal} {Phys. Rev. A}\ }\textbf {\bibinfo {volume} {77}},\
  \bibinfo {pages} {042108} (\bibinfo {year} {2008})}\BibitemShut {NoStop}%
\bibitem [{\citenamefont {Mitchison}\ \emph {et~al.}(2007)\citenamefont
  {Mitchison}, \citenamefont {Jozsa},\ and\ \citenamefont {Popescu}}]{SQM}%
  \BibitemOpen
  \bibfield  {author} {\bibinfo {author} {\bibfnamefont {G.}~\bibnamefont
  {Mitchison}}, \bibinfo {author} {\bibfnamefont {R.}~\bibnamefont {Jozsa}}, \
  and\ \bibinfo {author} {\bibfnamefont {S.}~\bibnamefont {Popescu}},\ }\href
  {\doibase 10.1103/PhysRevA.76.062105} {\bibfield  {journal} {\bibinfo
  {journal} {Phys. Rev. A}\ }\textbf {\bibinfo {volume} {76}},\ \bibinfo
  {pages} {062105} (\bibinfo {year} {2007})}\BibitemShut {NoStop}%
\bibitem [{\citenamefont {Kampen}(2007)}]{Kampen20071}%
  \BibitemOpen
  \bibfield  {author} {\bibinfo {author} {\bibfnamefont {N.~V.}\ \bibnamefont
  {Kampen}},\ }in\ \href {\doibase
  http://dx.doi.org/10.1016/B978-044452965-7/50004-0} {\emph {\bibinfo
  {booktitle} {Stochastic Processes in Physics and Chemistry (Third
  Edition)}}},\ \bibinfo {series and number} {North-Holland Personal Library},\
  \bibinfo {editor} {edited by\ \bibinfo {editor} {\bibfnamefont {N.~V.}\
  \bibnamefont {Kampen}}}\ (\bibinfo  {publisher} {Elsevier},\ \bibinfo
  {address} {Amsterdam},\ \bibinfo {year} {2007})\ \bibinfo {edition} {third
  edition}\ ed.,\ pp.\ \bibinfo {pages} {1 -- 29}\BibitemShut {NoStop}%
\bibitem [{\citenamefont {Puentes}\ \emph {et~al.}(2012)\citenamefont
  {Puentes}, \citenamefont {Hermosa},\ and\ \citenamefont {Torres}}]{JQV1}%
  \BibitemOpen
  \bibfield  {author} {\bibinfo {author} {\bibfnamefont {G.}~\bibnamefont
  {Puentes}}, \bibinfo {author} {\bibfnamefont {N.}~\bibnamefont {Hermosa}}, \
  and\ \bibinfo {author} {\bibfnamefont {J.~P.}\ \bibnamefont {Torres}},\
  }\href {\doibase 10.1103/PhysRevLett.109.040401} {\bibfield  {journal}
  {\bibinfo  {journal} {Phys. Rev. Lett.}\ }\textbf {\bibinfo {volume} {109}},\
  \bibinfo {pages} {040401} (\bibinfo {year} {2012})}\BibitemShut {NoStop}%
\bibitem [{\citenamefont {Kobayashi}\ \emph {et~al.}(2012)\citenamefont
  {Kobayashi}, \citenamefont {Puentes},\ and\ \citenamefont {Shikano}}]{jqv}%
  \BibitemOpen
  \bibfield  {author} {\bibinfo {author} {\bibfnamefont {H.}~\bibnamefont
  {Kobayashi}}, \bibinfo {author} {\bibfnamefont {G.}~\bibnamefont {Puentes}},
  \ and\ \bibinfo {author} {\bibfnamefont {Y.}~\bibnamefont {Shikano}},\ }\href
  {\doibase 10.1103/PhysRevA.86.053805} {\bibfield  {journal} {\bibinfo
  {journal} {Phys. Rev. A}\ }\textbf {\bibinfo {volume} {86}},\ \bibinfo
  {pages} {053805} (\bibinfo {year} {2012})}\BibitemShut {NoStop}%
\bibitem [{\citenamefont {Turek}\ \emph {et~al.}(2015)\citenamefont {Turek},
  \citenamefont {Kobayashi}, \citenamefont {Akutsu}, \citenamefont {Sun},\ and\
  \citenamefont {Shikano}}]{turek}%
  \BibitemOpen
  \bibfield  {author} {\bibinfo {author} {\bibfnamefont {Y.}~\bibnamefont
  {Turek}}, \bibinfo {author} {\bibfnamefont {H.}~\bibnamefont {Kobayashi}},
  \bibinfo {author} {\bibfnamefont {T.}~\bibnamefont {Akutsu}}, \bibinfo
  {author} {\bibfnamefont {C.-P.}\ \bibnamefont {Sun}}, \ and\ \bibinfo
  {author} {\bibfnamefont {Y.}~\bibnamefont {Shikano}},\ }\href
  {http://stacks.iop.org/1367-2630/17/i=8/a=083029} {\bibfield  {journal}
  {\bibinfo  {journal} {New Journal of Physics}\ }\textbf {\bibinfo {volume}
  {17}},\ \bibinfo {pages} {083029} (\bibinfo {year} {2015})}\BibitemShut
  {NoStop}%
\bibitem [{\citenamefont {Rendell}\ and\ \citenamefont
  {Rajagopal}(2005)}]{raja}%
  \BibitemOpen
  \bibfield  {author} {\bibinfo {author} {\bibfnamefont {R.~W.}\ \bibnamefont
  {Rendell}}\ and\ \bibinfo {author} {\bibfnamefont {A.~K.}\ \bibnamefont
  {Rajagopal}},\ }\href {\doibase 10.1103/PhysRevA.72.012330} {\bibfield
  {journal} {\bibinfo  {journal} {Phys. Rev. A}\ }\textbf {\bibinfo {volume}
  {72}},\ \bibinfo {pages} {012330} (\bibinfo {year} {2005})}\BibitemShut
  {NoStop}%
\bibitem [{\citenamefont {Adesso}\ \emph {et~al.}(2006)\citenamefont {Adesso},
  \citenamefont {Serafini},\ and\ \citenamefont {Illuminati}}]{gauss}%
  \BibitemOpen
  \bibfield  {author} {\bibinfo {author} {\bibfnamefont {G.}~\bibnamefont
  {Adesso}}, \bibinfo {author} {\bibfnamefont {A.}~\bibnamefont {Serafini}}, \
  and\ \bibinfo {author} {\bibfnamefont {F.}~\bibnamefont {Illuminati}},\
  }\href {\doibase 10.1103/PhysRevA.73.032345} {\bibfield  {journal} {\bibinfo
  {journal} {Phys. Rev. A}\ }\textbf {\bibinfo {volume} {73}},\ \bibinfo
  {pages} {032345} (\bibinfo {year} {2006})}\BibitemShut {NoStop}%
\bibitem [{\citenamefont {Simon}(2000)}]{simon}%
  \BibitemOpen
  \bibfield  {author} {\bibinfo {author} {\bibfnamefont {R.}~\bibnamefont
  {Simon}},\ }\href {\doibase 10.1103/PhysRevLett.84.2726} {\bibfield
  {journal} {\bibinfo  {journal} {Phys. Rev. Lett.}\ }\textbf {\bibinfo
  {volume} {84}},\ \bibinfo {pages} {2726} (\bibinfo {year}
  {2000})}\BibitemShut {NoStop}%
\bibitem [{\citenamefont {{Piacentini}}\ \emph {et~al.}()\citenamefont
  {{Piacentini}}, \citenamefont {{Levi}}, \citenamefont {{Avella}},
  \citenamefont {{Cohen}}, \citenamefont {{Lussana}}, \citenamefont {{Villa}},
  \citenamefont {{Tosi}}, \citenamefont {{Zappa}}, \citenamefont {{Gramegna}},
  \citenamefont {{Brida}}, \citenamefont {{Degiovanni}},\ and\ \citenamefont
  {{Genovese}}}]{marco}%
  \BibitemOpen
  \bibfield  {author} {\bibinfo {author} {\bibfnamefont {F.}~\bibnamefont
  {{Piacentini}}}, \bibinfo {author} {\bibfnamefont {M.~P.}\ \bibnamefont
  {{Levi}}}, \bibinfo {author} {\bibfnamefont {A.}~\bibnamefont {{Avella}}},
  \bibinfo {author} {\bibfnamefont {E.}~\bibnamefont {{Cohen}}}, \bibinfo
  {author} {\bibfnamefont {R.}~\bibnamefont {{Lussana}}}, \bibinfo {author}
  {\bibfnamefont {F.}~\bibnamefont {{Villa}}}, \bibinfo {author} {\bibfnamefont
  {A.}~\bibnamefont {{Tosi}}}, \bibinfo {author} {\bibfnamefont
  {F.}~\bibnamefont {{Zappa}}}, \bibinfo {author} {\bibfnamefont
  {M.}~\bibnamefont {{Gramegna}}}, \bibinfo {author} {\bibfnamefont
  {G.}~\bibnamefont {{Brida}}}, \bibinfo {author} {\bibfnamefont {I.~P.}\
  \bibnamefont {{Degiovanni}}}, \ and\ \bibinfo {author} {\bibfnamefont
  {M.}~\bibnamefont {{Genovese}}},\ }\href@noop {} {\ }\Eprint
  {http://arxiv.org/abs/1508.03220} {arXiv:1508.03220 [quant-ph]} \BibitemShut
  {NoStop}%
\end{thebibliography}%
 

\appendix
\section{}
In this section we illustrate how the correlation between pointer position variables gives rise to correlation between position and momentum variables. Here we assume the initial pointer state distribution as a Gaussian function represented by $\phi(q_{1},q_{2})$, whence the position distribution function $f(q_{1},q_{2})$ is given by
\begin{align}
f(q_{1},q_{2})& = |\phi(q_{1},q_{2})|^{2} \nonumber \\
\label{a1}
& = \sqrt{\frac{|G|}{(2\pi)^{2}}}exp[-\frac{1}{2}\sigma_{1}^{2}q_{1}^{2}-\frac{1}{2}\sigma_{2}^{2}q_{2}^{2}-corr(q_{1},q_{2})q_{1}q_{2}]
\end{align}
where $\langle \hat{q_{1}} \rangle_{in}=\langle \hat{q_{2}} \rangle_{in}=0$ and $|G|$ is the determinant of the covariance matrix written as
\begin{equation}
\label{a2}
G=\left[{\begin{array}{cc}
\sigma_{1} & corr(q_{1},q_{2})\\
corr(q_{1},q_{2}) & \sigma_{2} \end{array}}\right] 
\end{equation}

Now, note that for the above mentioned Gaussian  $f(q_{1},q_{2})$, non-zero correlation between $q_{1}$ and $q_{2}$ implies the non-separability of $f(q_{1},q_{2})$. Next, we show that the non-separability of $f(q_{1},q_{2})$ implies non-separability of $f(p_{1},q_{2})$ which in turn, entails non-zero correlation between $p_{1}$ and $q_{2}$. \\
From Eq. (\ref{a1}), taking the partial Fourier transformation from $q_{1}$ to $p_{1}$, one can obtain as follows (by ignoring the normalization constant)
\begin{align}
f(p_{1},q_{2})& = \int e^{-ip_{1}q_{1}}f(q_{1},q_{2})dq_{1} \nonumber \\
& = \int e^{-ip_{1}q_{1}}exp[-\frac{1}{2}\sigma_{1}^{2}q_{1}^{2}-\frac{1}{2}\sigma_{2}^{2}q_{2}^{2}-\nonumber \\
&   corr(q_{1},q_{2})q_{1}q_{2}]dq_{1} \nonumber \\
& = exp[-\frac{1}{2}(\sigma_{2}^{2}-\frac{corr^{2}(q_{1},q_{2})}{\sigma_{1}^{2}})q_{2}^{2}-\frac{p_{1}^{2}}{2\sigma_{1}^{2}}+\nonumber \\
\label{a3}
&   i\frac{corr(q_{1},q_{2})}{\sigma_{1}^{2}}p_{1}q_{2}]
\end{align}  
Then, using Eq. (\ref{sqm19.5}) in the text, one can evaluate the correlation between $\hat{p}_{1}$ and $\hat{q}_{2}$ for the function $f(p_{1},q_{2})$ given by Eq. (\ref{a3}). Thus we obtain 
\begin{align}
corr(p_{1},q_{2}) & = \int p_{1}q_{2}f(p_{1},q_{2})dp_{1}dq_{2}-\int p_{1}f(p_{1},q_{2})dp_{1}dq_{2} \nonumber \\
& \int q_{2}f(p_{1},q_{2})dp_{1}dq_{2} \nonumber \\
\label{a5}
& = i\frac{[corr(q_{1},q_{2})]}{\sigma_{1}^{2}} 
\end{align}
From Eq. (\ref{a5}) it is evident that if $corr(q_{1},q_{2}) \neq 0$, $corr(p_{1},q_{2})$ is also necessarily non-vanishing.
\end{document}